\documentclass[superscriptaddress,prl,floats,amsmath,amssymb,aps,twocolumn,showpacs]{revtex4-1}

\usepackage{graphicx}
\usepackage{amsmath}
\usepackage{amssymb}
\usepackage{dcolumn}
\usepackage{color}

\begin{document}
\title{A ferroelectric problem beyond the conventional scaling law}
\author{Qi-Jun Ye}
\affiliation{State Key Laboratory for Artificial Microstructure and Mesoscopic Physics, and School of Physics, Peking University, Beijing 100871, P. R. China}
\author{Zhi-Yuan Liu}
\affiliation{State Key Laboratory for Artificial Microstructure and Mesoscopic Physics, and School of Physics, Peking University, Beijing 100871, P. R. China}
\author{Yexin Feng}
\affiliation{School of Physics and Electronics, Hunan University, Changsha 410082, P. R. China}
\author{Peng Gao}
\affiliation{Electron Microscopy Laboratory, International Center for Quantum Materials, and
School of Physics, Peking University, Beijing 100871, P. R. China}
\affiliation{Collaborative Innovation Center of Quantum Matter, Peking University, Beijing 100871, P. R. China}
\author{Xin-Zheng Li}
\email{xzli@pku.edu.cn}
\affiliation{State Key Laboratory for Artificial Microstructure and Mesoscopic Physics, and School of Physics, Peking University, Beijing 100871, P. R. China}
\affiliation{Collaborative Innovation Center of Quantum Matter, Peking University, Beijing 100871, P. R. China}
\date{\today}
\begin{abstract}
Ferroelectric (FE) size effects against the scaling law were reported recently
in ultrathin group-IV monochalcogenides, and extrinsic effects (e.g. defects and lattice
strains) were often resorted to.
Via first-principles based finite-temperature ($T$) simulations,
we reveal that these abnormalities are intrinsic to their unusual symmetry breaking
from bulk to thin film.
Changes of the electronic structures result in different order parameters characterizing the FE phase
transition in bulk and in thin films, and invalidation of the scaling law.
Beyond the scaling law $T_{\text{c}}$ limit, this mechanism can help predicting materials promising
for room-$T$ ultrathin FE devices of broad interest.
\end{abstract}


%
%
%

\maketitle

\clearpage

\begin{figure}
\includegraphics[width=0.88\linewidth]{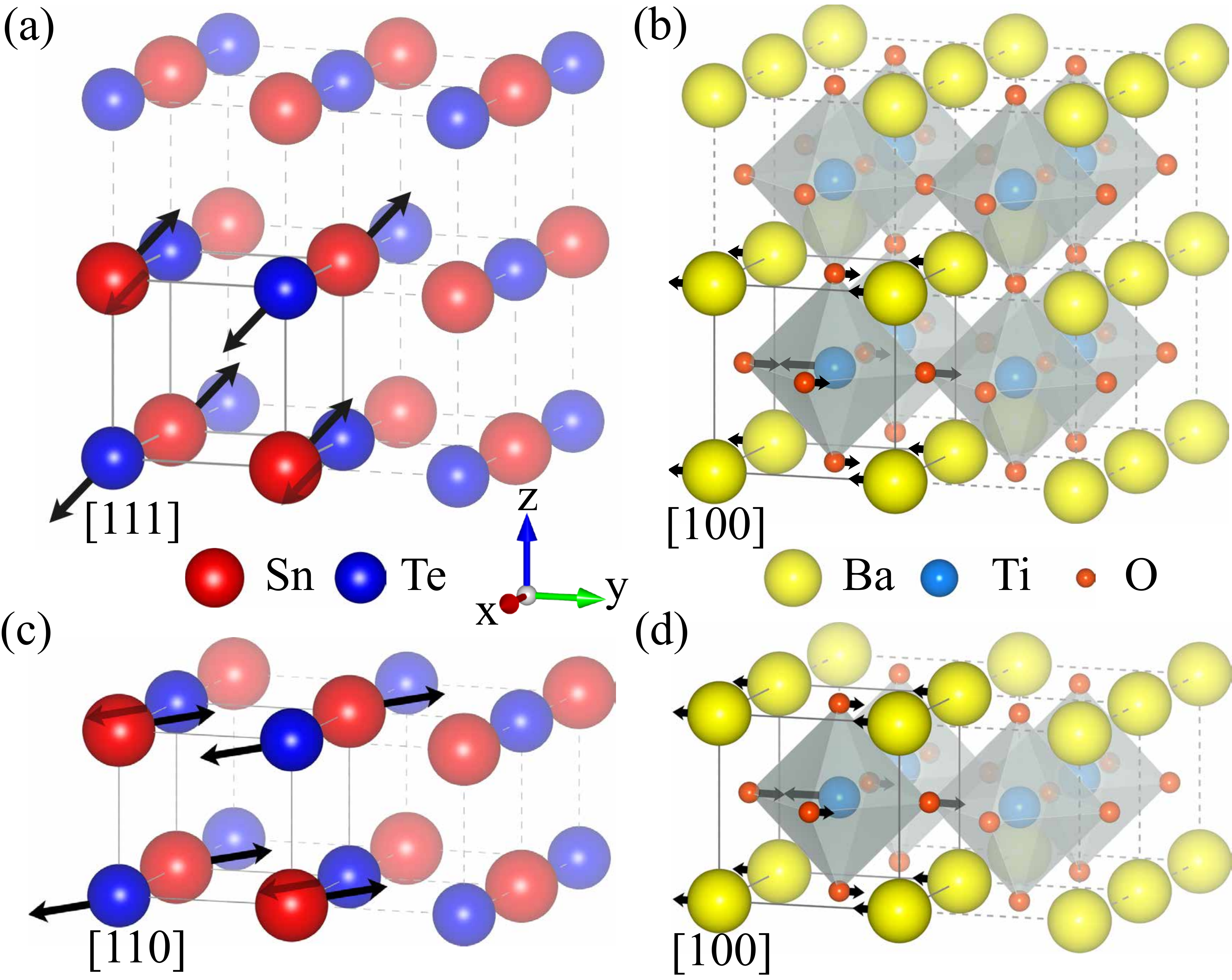}
\caption{\label{fig1} The responsible FE modes associated with the FE phase in (a) bulk SnTe, (b) 1UC SnTe film,
(c) bulk BTO, and (d) 1UC BTO film (Ba-O terminated). The black arrows indicates atomic displacements along the FE modes.}
\end{figure}

\begin{figure}
\includegraphics[width=0.93\linewidth]{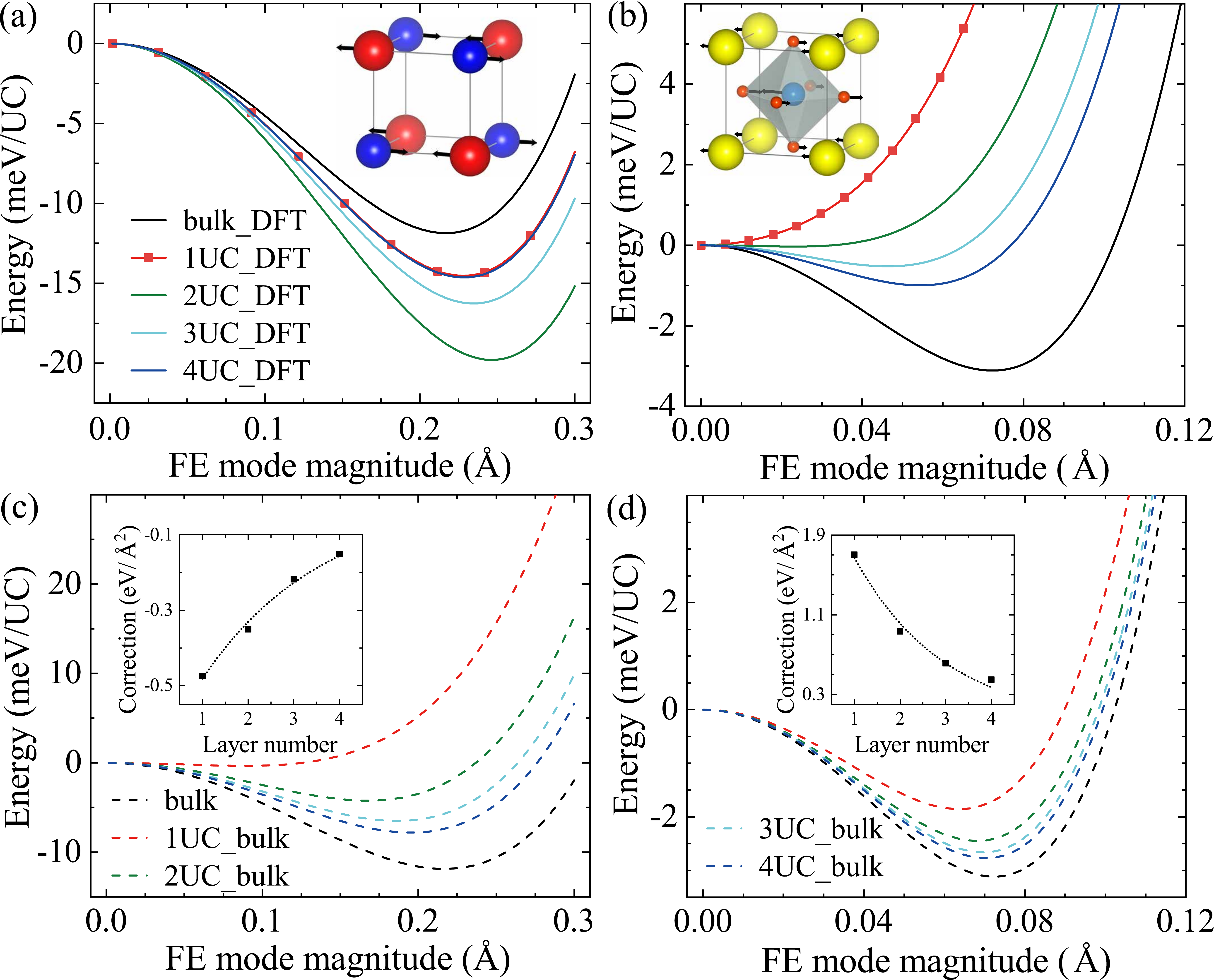}
\caption{\label{fig2} The potential energy curves of bulk and 1-4UC thin films by DFT (solid lines) for (a) SnTe and (b) BTO. Bottom panels show the same curves using Eq.~(\ref{totenergy}) only with the first two
terms for (c) SnTe and (d) BTO. The insets in (c) and (d) show the magnitude of the quadratic
corrections in Eq.~(\ref{2D_corr}).}
\end{figure}

Miniaturized ferroelectric (FE) device of continued demand in portable consumer electronics
poses prerequisite understandings of a fundamental question, \textit{i.e.} the nature of FE size effects
~\cite{Junquera2003,Spaldin2004,Ahn2004,Dawber2005,Scott2007,Martin2016}.
Finite size scaling (FSS) theory, as the conventional wisdom, predicts that the Curie
temperature $T_\text{c}$
for the paraelectric (PE) to FE phase transitions decreases when scaling down to finite sizes~\cite{Fisher1972,Binder1974,Huang1993}, following:
\begin{equation}
\delta T_\text{c}(d)=\frac{T_\text{c}(\infty)-T_\text{c}(d)}{T_\text{c}(\infty)}
=\left(\frac{\xi_0}{d}\right)^\lambda,
\end{equation}
where $T_\text{c}(d)$ and $T_\text{c}(\infty)$ are the $T_\text{c}$ of the film of thickness $d$ and
bulk, respectively~\footnote{The scaling law is sometimes written in a similar form suggested
to better fit the experimental data, as $\delta T^\prime_\text{c}(d)=(T_\text{c}(\infty)-T_\text{c}(d))/T_\text{c}(d)
=\left(\xi_0^\prime/d\right)^{\lambda^{\prime}}$.}.
The $T_{\text{c}}$s of different sizes are related via the character length $\xi_0$ and
the universal critical exponent $\lambda$.
As FSS theory shown predictive in perovskite compounds and a variety of FEs~\cite{Junquera2003,Fong2004,Kornev2007,Almahmoud2010},
$T_\text{c}(d)$ being lower in ultrathin films was believed heretofore
as an essential limit in realizing room temperature ($T$) ultrathin FE devices of broad
interest~\cite{Spaldin2004,Fong2004}.
Recent studies on group-IV monochalcogenides, however, opened the door for realization
of room $T$ ultrathin FE devices beyond the FSS theory prediction
~\cite{Chang2016,Fei2016,Wu2016,Mehboudi2016,Wan2017,Liu2018}.
The experiment by K. Chang \textit{et al.} showed that in one unit-cell (1UC) SnTe film the
Curie temperature ($T_{\text{c}}^{\text{1UC}}$) is 270 K~\cite{Chang2016}, enhanced from the bulk
value ($T_\text{c}^{\text{bulk}}$) of 98~K~\cite{Iizumi1975}.
%
Parallel to this, Fei \textit{et al.} predicted robust ferroelectricity in analogous monolayer
group-IV monochalcogenides MX (M = Ge, Sn; X = S, Se) via the Landau-Ginzburg type effective
Hamiltonian method~\cite{Fei2016}.
Wu and Zeng showed MX's multiferroelectricity, where the polarization valley switching by using
stress or electric field enables designing room-$T$ nonvolatile memory~\cite{Wu2016,Hanakata2016}.
Nevertheless, large extrinsic effects claimed in these studies such as lower free carrier 
density~\cite{Sugai1977,Kobayashi1976,Chang2016,Mehboudi2016}, lattice strains~\cite{DeWette1985,Samara1975,Wan2017},
etc. render the intrinsic size effect of ferroelectricity unimportant, thereby hindering further investigation and searching for other promising materials.
In this letter, we address two issues:
i) reveal the nature of intrinsic FE size effects in these materials and analyze their relation with
the FSS theory;
ii) propose an easy-to-use criteria for potential low-dimensional FE materials with $T_\text{c}$ higher
than their high-dimensional correspondences.
SnTe and BaTiO$_3$ (BTO), two paradigmatic FE materials whose scaling behaviors show remarkable difference,
are discussed in details.
Based on the first-principles exploration of potential energy surfaces, an effective Hamiltonian is built
and used in Monte Carlo simulations, to investigate the finite-$T$ PE-FE phase transitions.
Our simulations reproduce the experimental results of robust in-plane ferroelectricity and
abnormal thickness dependency of the $T_{\text{c}}$ in SnTe films, and the conventional scaling
behaviors in BTO films.
The key factor, we identify in SnTe the order parameters are deviated for the 3D and 2D PE-FE
phase transitions, while in BTO no deviation occurs, is essential to understand this fundamental difference.
As this can be perceived macroscopically by jumping phases in the PE-FE transition, a rule of thumb
is proposed to predict analogous low-dimensional FE materials.
We adopt the model proposed by Vanderbilt and coworkers~\cite{Zhong1995,Bellaiche2000a}, which enables
large-scale calculations with first-principles predictive power, to investigate the FE phase transitions
in bulk and thin films of SnTe and BTO.
This method is formerly and successfully applied to bulk perovskites including BTO
~\cite{Bellaiche2000b,Kornev2004,Meyer2002,Nishimatsu2008,Kornev2005,Chen2015}.
Due to the same displacive feature, \textit{i.e.} soft optical modes (so-called FE modes) driving spontaneous
polarization below $T_{\text{c}}$, we develop this method for group-IV monochalcogenides
including SnTe, and for their thin films.
The total energy of an instantaneous finite-$T$ structure differing from the reference perfect
crystal state is written as
\begin{equation}
\label{totenergy}
E_{\text{tot}}^{(d_{\text{m}})}=E_{\text{ref}}+E_{\text{3D-param}}^{(d_\text{m})}\left(\{\mathbf{u}_i\},\eta,p\right) +E_{\text{corr}}^{(d_{\text{m}})},
\end{equation}
where $d_{\text{m}}$ labels the dimension of the system, $\mathbf{u}_i$ describes the FE modes
at $i$-th site, $\eta$ is the homogeneous strain tensor, and $p$ is the hydrostatic pressure
coupled with the diagonal terms of $\eta$.
$E_{\text{3D-param}}^{d_\text{m}}$ contains the intra- and interactions of the dominant soft modes
(FE modes here) and the lattice strains, parameterized in 3D structure.
The specific form of these terms can be found in Refs.~[\onlinecite{Zhong1995}] and [\onlinecite{SI}], and
a schematic of one finite-$T$ instantaneous FE mode configuration on the strained lattice is shown in Fig.~S1.
The correction term $E_{\text{corr}}^{(d_{\text{m}})}$ is added only for the 2D and 1D systems to
address the changes of the electronic structures upon decreasing dimensionality, as we will show
later in Fig.~\ref{fig2}.
For ultrathin films (2D systems), we adopt a correction of exponential decay on the film thickness,
analogous to the form of Ref.~[\onlinecite{Almahmoud2004}], as
\begin{equation}
\label{2D_corr}
E_{\text{corr}}^{(\text{2D})}\left(n_l\right)=\sum_{\substack{\alpha=\beta\\ \alpha=x,y}}\sum_{\substack{\left<i,j\right>\\j=i\pm\hat{\alpha}}} e^{-B\cdot n_l} A_{ij,\alpha\beta}u_{i\alpha}u_{j\beta},
\end{equation}
where $n_l$ labels the number of layers, $A_{ij,\alpha\beta}$ describes the short range interactions
(exclude the short part of dipole-dipole interactions) between neighboring sites $\left<i,j\right>$.
Parameters for Eqs.~(\ref{totenergy}) and~(\ref{2D_corr}) are derived from first-principles
explorations of the potential energy profiles of the 3D and 2D systems, respectively.
More computation details please see our Supplementary Information~\cite{SI}.
FE modes, the key instabilities for system going from high-symmetry PE phase to symmetry-breaking FE phase,
can be viewed as the order parameter in this process.
In fact, it is a good approximation shown by Refs.~[\onlinecite{King-Smith1993}] and
[\onlinecite{Resta1993}] that the polarization at one unit site ($\mathbf{P}_i$) is almost linear to
the FE mode magnitude, through:
\begin{equation}
\mathbf{P}_i=eZ^*_{\text{Born}}\mathbf{u}_i/V.
\end{equation}
$Z^*_{\text{Born}}$ is the Born charge and $V$ is the cell volume.
We use ${u}_{x,y,z}=\left<{u}_i\right>_{x,y,z}$ to characterize the phase transition.
The responsible FE modes for the 3D and 2D structures are different in SnTe, and they are
the same in BTO.
Polarization along [111] in 3D SnTe [Fig.~\ref{fig1}(a)], namely the rhombohedral FE phase, is a
results of simultaneous softening of the triply degenerate FE modes ${u}_{x}$, ${u}_{y}$, and ${u}_{z}$.
Whilst in 2D SnTe [Fig.~\ref{fig1}(c)], it is the polarization along [110] and the softening of the
doubly degenerate in-plane FE modes ${u}_{x}$ and ${u}_{y}$ which characterize the PE-FE transition.
In BTO [Figs.~\ref{fig1}(b) and \ref{fig1}(d)], the polarization along [100] and the softening of a
singlet FE mode is the order parameter, and it does not change in the 3D and
2D systems\footnote{Here we talk about the Curie temperature related FE phase of BTO,
namely the tetragonal phase. There are two other FE phases, where [110] and [111] polarizations
require more FE modes involved}.
This unusual symmetry breaking in SnTe might be a clue to its abnormal scaling behavior.

We start discussions by looking at the static energies.
Taking the cubic structure as reference, we arrange the Sn and Te atoms (Ba, Ti, and O atoms for BTO) following the displacement patterns of the soft modes and monitor the total energy variations.
Figs.~\ref{fig2}(a) and \ref{fig2}(b) show the DFT potential curves along one FE mode in the bulk and
the 1-4UC films of SnTe and BTO, respectively.
The bulk results are approached in both two materials upon increasing the film layers, whereas different
evolutions are observed.
In 1-4UC films of SnTe, the deeper potential wells permit larger instabilities for soft modes, implying
an enhancement of $T_{\text{c}}$ in the films.
Moreover, the abnormal weakening of this soften feature in the 1UC film compared with the 2-4UC films,
suggests a non-monotonous variation of the $T_{\text{c}}$ in 1-4UC films.
In BTO, the FE soft mode is monotonously weakened in the films, implying a conventional scaling behavior.
The dashed lines in Figs.~\ref{fig2}(c) and \ref{fig2}(d) are results obtained using only the first two terms
in Eq.~(\ref{totenergy}), shown to highlight the importance of $E_{\text{corr}}^{(d_{\text{m}})}$ in Eq.~(\ref{2D_corr}).
Without the correction term $E_{\text{corr}}^{(d_{\text{m}})}$, the total for SnTe in Eq.~(\ref{totenergy})
is clearly off the trend of DFT curves [Fig.~\ref{fig2}(c)].
For BTO it differs quantitatively [Fig.~\ref{fig2}(d)].
$E_{\text{corr}}^{(d_{\text{m}})}$ represents the intrinsic changes of the electronic structures
upon changing from bulk to thin films~\cite{Liu2018}.
Its magnitude as a function of layers is shown in the inset of Figs.~\ref{fig2}(c) and \ref{fig2}(d).
The different roles played by $E_{\text{corr}}^{(d_{\text{m}})}$ in SnTe and BTO is
crucial for their scaling behaviors.
These static DFT results are in alignment with the experiments in Ref.~\cite{Chang2016}.
However, considering the complicated 2D nature, they are not sufficient to
clarify the full picture of the FE phase transitions in thin films at finite-$T$s.

\begin{figure}[t]
\includegraphics[width=0.95\linewidth]{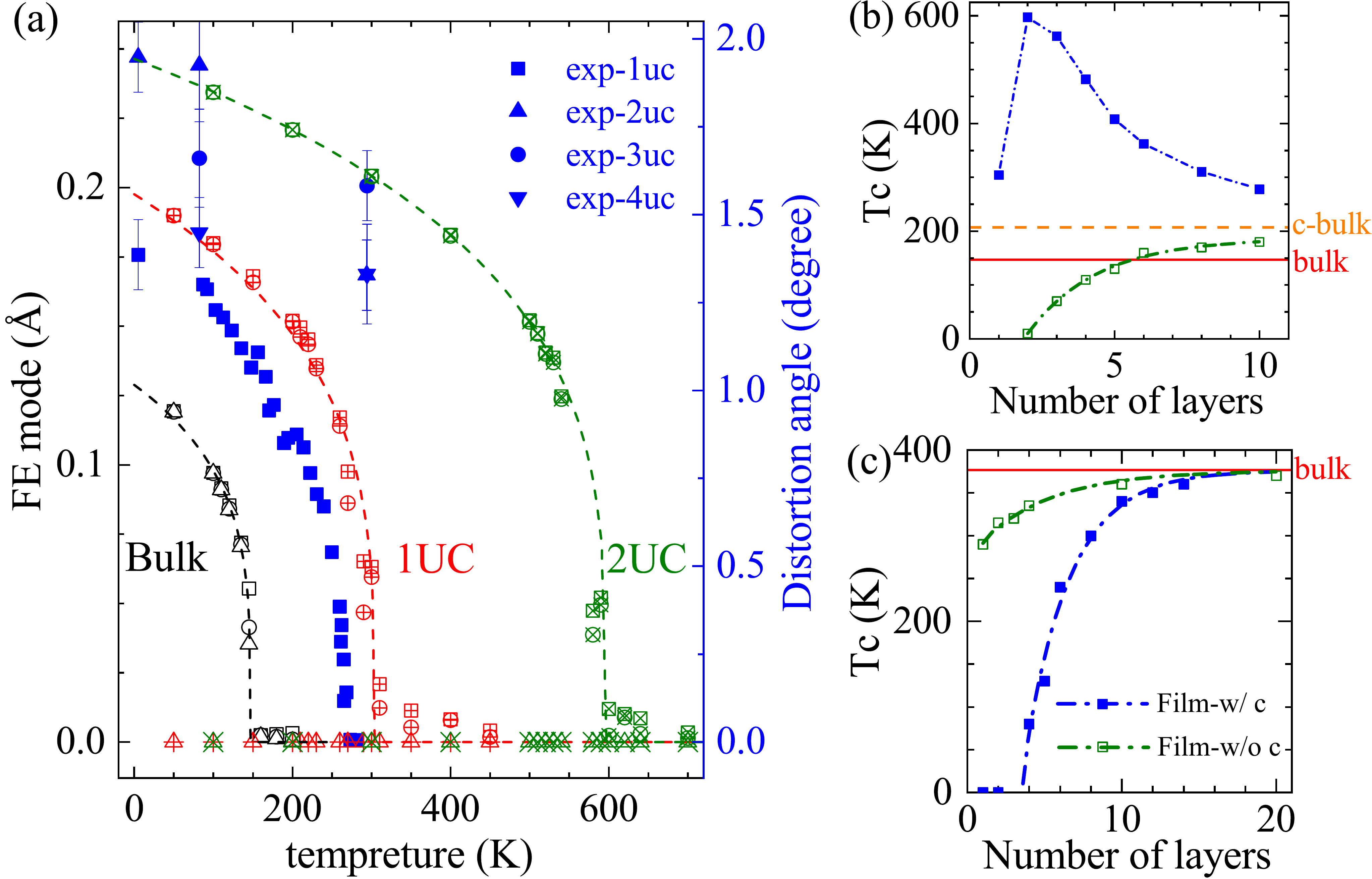}
\caption{\label{fig3}(a) The phase transitions in bulk, 1UC, and 2UC thin films of SnTe (in black, red, and green
open marks, respectively). The order parameters, $u_x$, $u_y$, and $u_z$ are characterized by 
square, sphere, and triangle, respectively. 
In bulk, $u_x=u_y=u_z\neq0$ in the FE phase. In films, $u_x=u_y\neq0$ and $u_z=0$ in the FE phase.
Blue solid marks show the experimental data acquired from Ref.~[\onlinecite{Chang2016}].
Thickness dependency of $T_{\text{c}}$ is show in (b) for SnTe and in (c) for BaTiO$_3$.
Blue (olive) curves show the case with (without) considering $E_{\text{corr}}^{(d_{\text{m}})}$.
}
\end{figure}

\begin{figure}
\includegraphics[width=0.88\linewidth]{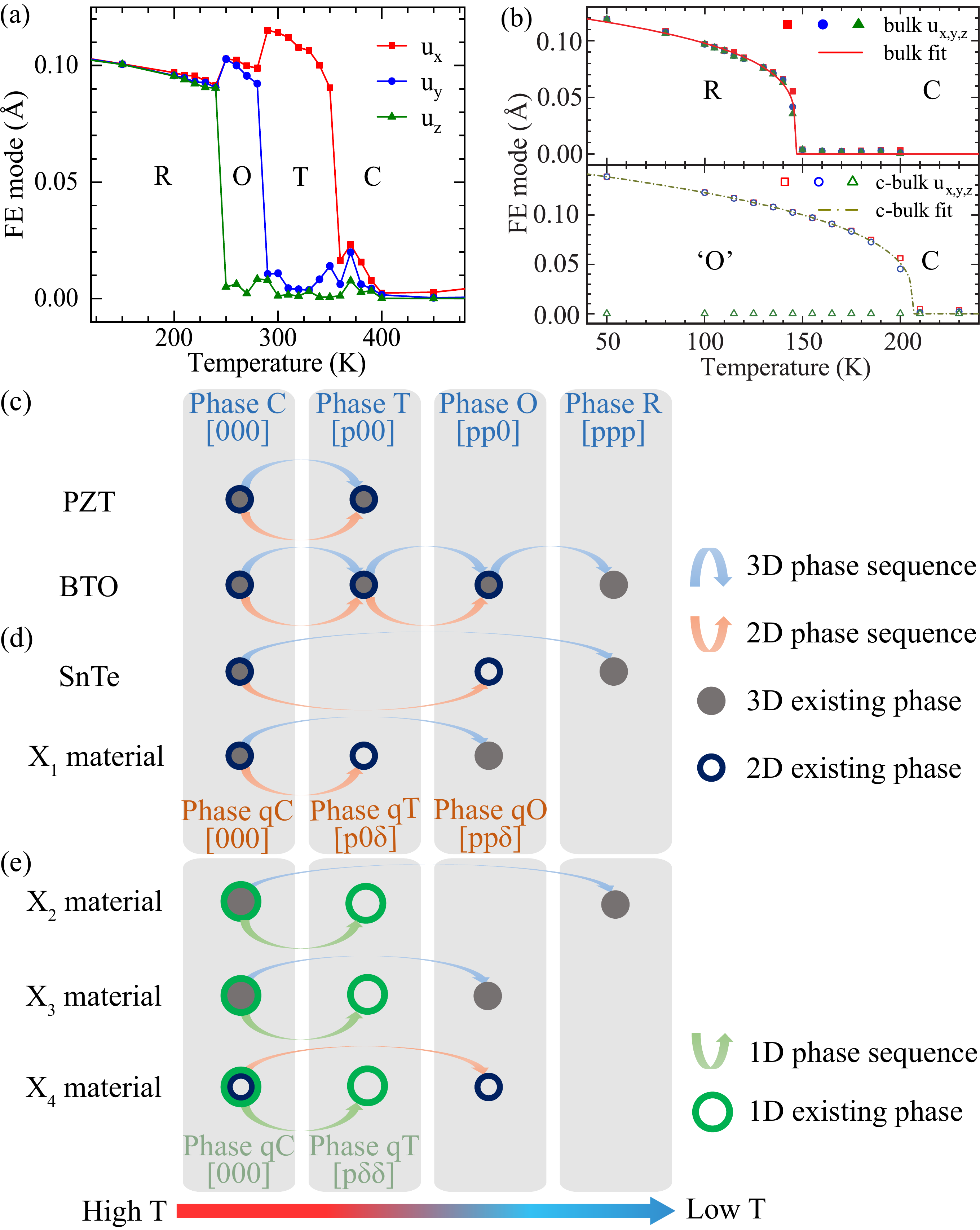}
\caption{\label{fig4}(a) the C-T-O-R transition sequence in bulk BTO. (b) top panel: the spontaneous C-R phase transition in bulk SnTe; bottom panel: the artificial C-O phase transition in bulk SnTe by constraining $u_{z}=0$. (c)-(e) Schematic of the transition sequence in 3D and 2D FE materials. (c) the conventional ones including BTO and PZT; (d) 2D anomaly including SnTe;
(e) 1D anomaly based on the same mechanism remains to be explored. X$_1$ to X$_4$ label candidates for
robust low-dimensional FE devices.}
\end{figure}

To obtain $T_{\text{c}}$, we use the aforementioned effective Hamiltonian to perform finite-$T$
Monte-Carlo simulations.
We first look at the bulk PE-FE phase transitions in SnTe and BTO.
SnTe turns from cubic PE phase (Fm-3m) to rhombohedral (R3m) FE phase at 98 K~\cite{Iizumi1975}.
Our simulations reproduce this by giving a $T_{\text{c}}$ of 147 K, which is
identified by the temperature dependency of FE modes [black marks in Fig.~\ref{fig3}(a)].
Difference of $\sim$50 K is left to account for defects effect, which is absent in our
perfect crystal simulations~\cite{Sugai1977,Kobayashi1976}.
Our simulations also obtain reasonable $T_{\text{c}}\sim370$ K for bulk BTO transiting from cubic PE
phase (Pm-3m) to tetragonal FE phase (P4mm) [Fig.~S8], consistent with published studies~\cite{Zhong1994,Walizer2006}.
Then we check $T_{\text{c}}$ at varying layers.
Deviated from bulk, the SnTe monolayer prefers in-plane polarization (along $\left[110\right]$ 
direction)~\cite{Huang2003}, as shown by red marks in Fig.~\ref{fig3}(a).
We observed a transition from the PE tetragonal phase to the FE monoclinic phase.
This in-plane polarization in monolayer is robust even at room $T$, appealing for practical ultrathin devices.
Besides this, the thickness dependency of $T_\text{c}$ is also in alignment with the
experimental observations, which measures the distortion angles [Fig.~\ref{fig3}(a) scale to right in blue].
This can be seen by comparing the trend of saturated distortion angle with FE modes from our simulations.
They are smaller in 1UC than in 2UC [from red to green symbols scaling to left in Fig.~\ref{fig3}(a)].
After 2UC, they decreases and approaches the bulk value from above.
More alignments can be found in the magnitude of saturated distortion 
angle $\sim1.2^\circ$ (exp. $\sim1.4^\circ$),
and the critical index 0.27-0.35 for 1-4UC films (exp. 0.33$\pm$0.05), see our SI \cite{SI}.
$T_{\text{c}}$ shows the same non-monotonous trend in clear discrepancy with the
conventional scaling law [blue curves in Fig.~\ref{fig3}(b)], whereas it holds in BTO [Fig.~\ref{fig3}(c)].
Last but not least, threats from extrinsic effects should be ruled out or controlled.
Considering the fact we reproduce the abnormal scaling behavior of SnTe upon using stoichiometric
structure, the effects of free carriers (Sn vacancies) should be minor~\cite{Sugai1977,Kobayashi1976}.
Strain effects, however, are crucial and might dramatically tune $T_{\text{c}}$ shown by early studies
in perovskites~\cite{Haeni2004,Choi2004,Fei2016}.
Since our model exhibits a build-in stress-strain relation, we set the same external pressure 
and fully relax the films in the MC simulations.
In so doing, we claim the abnormalities in SnTe is an intrinsic size effect with underlying mechanism to be revealed.
To understand this abnormality, we compare the microscopic details of the PE-FE phase transitions
in SnTe and in BTO.
In bulk BTO, four phases from cubic (C) through tetragonal (T) and orthogonal (O) to
rhombohedral (R) exist upon decreasing $T$s, and polarizations along $x$, $y$, and $z$ appear
sequentially [Fig.~\ref{fig4}(a)].
In BTO thin films, depolarization results in zero polarization along $z$.
Three phases from quasi-cubic (qC) through quasi-tetragonal (qT) to
quasi-orthogonal (qO) exist at decreasing $T$s, and polarizations along $x$ and $y$
appear sequentially.
In both cases, $T_{\text{c}}$ corresponds to the same physical process (symmetry breaking here) that only
one of the three FE modes is soften [Fig.~\ref{fig4}(a)], \textit{i.e.} C-T phase transition in
bulk and qC-qT one in thin films.
From bulk to thin films, the finite film thickness cuts off long-distance correlations along $z$
of the in-plane polarizations so that an appreciable finite-size rounding of critical-point
singularities is to be expected~\cite{Privman1990}.
This forms the basis of FSS theory~\cite{Fisher1972}, and conventional scaling behavior is
expected.
This situation, however, is different in SnTe where $T_{\text{c}}$ corresponds to different
physical processes, as discussed.
The PE-FE transition is C-R in bulk and qC-qO in films.
The qC-qO transition in films corresponds to the C-O transition in bulk, which does not
appear spontaneously.
Utilizing the knowledge of BTO's phase sequence, if C-O exists, it should occur at a higher $T$.
By convenience of our simulations, we can verify this by constraining the FE mode along $z$ direction $u_z=0$.
This allows us to artificially obtain the C-O transition sequence in bulk, as shown in Fig.~\ref{fig4}(b).
When the PE-FE transition is forced to happen between C and O phases, the $T_{\text{c}}$ is substantially
elevated.
Therefore, the elevated $T_{\text{c}}$ in the films is related to this omitted O phase in bulk.
The FSS theory aims to describe the scaling behavior between universality classes only deviated in spatial
dimensionality, which presumes the same physical process, characterized by the same order parameters and
formulation of interactions upon scaling the system size.
This prerequisite is not fulfilled in SnTe.
The order parameters clearly change since the triply degenerate FE modes can not soften simultaneously in the films.

These different scaling behaviors can also be understood by looking at the role played by
$E_{\text{corr}}^{(d_{\text{m}})}$ in Eq.~(\ref{totenergy}).
In BTO, the quantitative changes in electronic structure do not result in a qualitative
change of their relative positions upon going from the films to bulk.
Whilst this is not the case in SnTe [Figs.~\ref{fig2}(a) and~\ref{fig2}(c)].
In Refs.~[\onlinecite{Fisher1972}], ~[\onlinecite{Fisher1974}], and [\onlinecite{Wilson1983}], 
when the scaling is deduced, a model Hamiltonian (e.g. the Ising model) is chosen and the 
difference between the bulk and films is characterized by geometric changes.
Renormalization group theory is used and the subtle but crucial changes of the Hamiltonian upon
going from bulk to films are neglected.
This assumption is violated seriously in SnTe.
To test this, we can choose the first two terms in Eq.~(\ref{totenergy}), which addressed the geometric
changes but not the electronic structures, to perform the PE-FE phase transition upon going from
bulk to films.
The scaling law becomes valid again in both SnTe and BTO [olive curves in Figs.~\ref{fig3}(b)
and~\ref{fig3}(c)].
Therefore, when the changes of electronic structures result in a qualitative change of the
Hamiltonian itself, the scaling law fails.
One macroscopic observable to characterize this abnormality is the order parameters related to symmetry
as we have discussed.
Using this picture, we now propose some promising low-dimensional FE materials with
higher $T_{\text{c}}$ than their higher-dimensional correspondences.
Jumping transition sequence in Figs.~\ref{fig4}(d)-(e) could help.
Intuitively, this means highly degenerate FE modes, which can soften simultaneously in the
higher dimensional systems.
With decreasing dimensionality, symmetry breaking eliminates this simultaneous softening.
Thereby, one can expect different order parameters for bulk and films, and higher $T_{\text{c}}$
beyond the scaling law limit.
In bulk, a C-T-O-R sequence of phase transition might happen upon decreasing $T$s.
This corresponds to a qC-qT-qO sequence in films, and a qC-qT sequence in 1D systems.
Upon going from 3D to 2D, Fig.~\ref{fig4}(c) shows the case when nothing was jumped in bulk, including
BTO and Pb[Zr$_x$Ti$_{1-x}$]O$_3$ (PZT).
In SnTe, the T \& O phases were jumped.
Besides this, when the 3D PE-FE phase transition happens between C \& O, the T phase can be
jumped [X$_1$ in Fig.~\ref{fig4}(d)].
This picture might also apply to the 3D to 1D and 2D to 1D transitions.
Three possibilities are shown in Fig.~\ref{fig4}(e).
When the 3D PE-FE transition happens between C \& R (or C \& O), the T \& O phases (T phase) are
jumped in bulk, labeled by X$_2$ (X$_3$).
When the 2D PE-FE transition happens between qC \& qR, the qT phase is jumped in the
films [X$_4$ in Fig.~\ref{fig4}(e)].
These suggestions based on symmetry provide a simple rule of thumb to seek systems in which
the low-dimensional systems can possess higher $T_{\text{c}}$ than their higher-dimensional correspondences.
Accurate numerical characterizations, however, need to resort to the first-principles
based finite-$T$ simulations as reported above.
Considering the fundamental importance of FE size effect and phase transition problems in condensed matter
physics, we hope this work can stimulate more experimental and theoretical studies in this direction.\\
\begin{acknowledgements}
The authors are supported by the National Basic Research Programs of China
under Grand Nos. 2016YFA0300900, the National Science Foundation of China under Grant Nos
11774004, 11604092, and 11634001. We sincerely thank Prof. Zhirong Liu and Prof. Wenhui
Duan for insightful discussions.
The computational resources were supported by the High-performance Computing Platform
of Peking University, China.
\end{acknowledgements}
%
%

\end{document}


\title{Supplementary Information:\\ A ferroelectric problem beyond the conventional scaling law}
\author{Qi-Jun Ye}
\affiliation{State Key Laboratory for Artificial Microstructure and Mesoscopic Physics, and School of Physics, Peking University, Beijing 100871, P. R. China}
\author{Zhi-Yuan Liu}
\affiliation{State Key Laboratory for Artificial Microstructure and Mesoscopic Physics, and School of Physics, Peking University, Beijing 100871, P. R. China}
\author{Yexin Feng}
\affiliation{School of Physics and Electronics, Hunan University, Changsha 410082, P. R. China}
\author{Peng Gao}
\affiliation{Electron Microscopy Laboratory, International Center for Quantum Materials, and
School of Physics, Peking University, Beijing 100871, P. R. China}
\affiliation{Collaborative Innovation Center of Quantum Matter, Peking University, Beijing 100871, P. R. China}
\author{Xin-Zheng Li}
\email{xzli@pku.edu.cn}
\affiliation{State Key Laboratory for Artificial Microstructure and Mesoscopic Physics, and School of Physics, Peking University, Beijing 100871, P. R. China}
\affiliation{Collaborative Innovation Center of Quantum Matter, Peking University, Beijing 100871, P. R. China}
\date{\today}
\maketitle

Here we provide more computational details of the simulations in our manuscript and additional discussions. In section I, the constructions of the effective Hamiltonian are shown for both 3D case and 2D case. The setups of DFT calculations and Monte Carlo simulations, and parameters for SnTe and BTO are given in Section II. Section III discusses the compensation pressure which is used to rule out the strain effect. Microscopic details of the phase transitions (such as the magnitude of FE mode, lattice strain) of SnTe bulk and thin films are given in Section IV and Section V, respectively. A pressure-temperature phase diagram for SnTe bulk is also given in Section IV. At last in Section VI, we process a numerical experiment in which case the scaling law is valid, to further support our discussions.

\section{Effective Hamiltonian method}
\subsection{3D case}
\begin{figure}[h]
\includegraphics[width=0.7\linewidth]{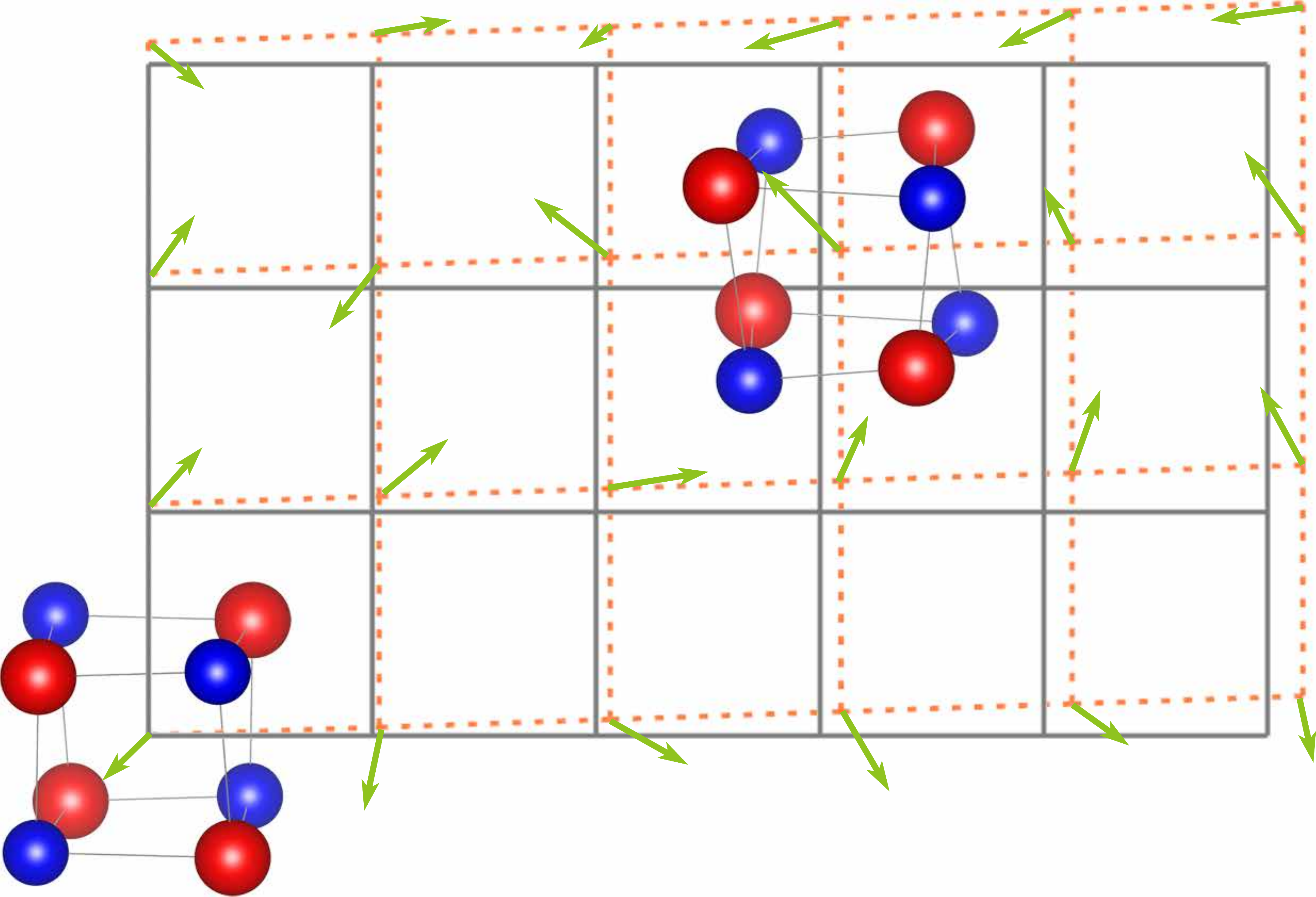}
\caption{\label{figs1}A schematic of the effective potential. The reference structure (grey solid lines) is distorted by lattice strain $\eta$ (orange dashed lines) and FE modes $\mathbf{u}_i$ \textit{i.e.} the polarization (green arrows). Their intra- and interactions are used later to perform the Monte-Carlo simulations.}
\end{figure}

We adopt the form of effective Hamiltonian in Ref.~[\onlinecite{Zhong1994}].
The effective Hamiltonian consists of three parts, as
\begin{equation}
E_{\text{3D-param}}^{(\text{3D})}\left(\{\mathbf{u}_i\},\eta,p\right)=E_1(\left\{\bm{u}_i\right\})+E_2(\left\{\eta_l\right\},p_{\text{ext}})+E_3(\left\{\bm{u}_i\right\},\left\{\eta_l\right\}),
\end{equation}
where $E_1(\left\{\bm{u}_i\right\})$ is the energy term of FE modes $\left\{\bm{u}_i,~ i=1,2,\cdots,N\right\}$, $E_2(\left\{\eta_l\right\},p_{\text{ext}})$ is the energy term of lattice strain $\left\{\eta_l,~l=1,2,\cdots,6\right\}$ and external pressure $p_{\text{ext}}$, and $E_3(\left\{\bm{u}_i\right\},\left\{\eta_l\right\})$ is their coupling.
The energy term of FE modes $E_1(\left\{\bm{u}_i\right\})$, contains the local and non-local terms of the FE modes, as
\begin{equation}
E_1(\left\{\bm{u}_i\right\})=E_{\text{self}}(\left\{\bm{u}_i\right\})+E^{(\text{3D})}_{\text{dipole}}(\left\{\bm{u}_i\right\})+E_{\text{short}}(\left\{\bm{u}_i\right\}),
\end{equation}
where $E_{\text{self}}$ is the isolated on-site energy of the FE modes, $E^{(\text{3D})}_{\text{dipole}}$ and $E_{\text{short}}$ describe their non-local interactions of long-range (here is the dipole-dipole interactions) and short-range (electron hybridization and repulsion between 1st-3rd nearest neighbors), respectively.
When atoms of one cell are separately displaced from their position of the perfect cubic structure, \textit{i.e.} atoms in the other cells keep unchanged, it yields the on-site energy of the FE modes.
We take the Taylor series to describe this energy variation, with cutoff up to fourth order as
\begin{equation}
\label{eself}
E_{\text{self}}=\sum_i\left(\kappa_2 u_i^2+\alpha_4 u_i^4+\sum_{(\alpha,\beta)}\gamma_4 u_{i\alpha}^2u_{i\beta}^2\right),
\end{equation}
where $\kappa_2$, $\alpha_4$, and $\gamma_4$ are the coefficients for corresponding orders.
The odd terms have been ignored according to cubic symmetry.
Since FE modes are soft modes, the fourth order terms are considered to contain at least anharmonic contributions.
For non-local parts, the most prominent term is the dipole-dipole interaction $E^{(\text{3D})}_{\text{dipole}}$.
When considering the macroscopic FE properties, it's a good approximation to view the polarization (distributed continuously in the space) as a series of departed point dipoles located at the center of each cell.
However, short-range repulsion and electron hybridization, \textit{i.e.} $E_{\text{short}}$, must be also included to complement the non-local interactions.
Analogous with the treatment of Coulomb terms and exchange-correlation terms in DFT, we shall exactly characterize the  dipole-dipole interaction in $E_{\text{dipole}}^{(\text{3D})}$ and include all the other inter-site terms in $E_{\text{short}}$.
We implement the $E_{\text{dipole}}$ calculations with Ewald summation (EW3D)\cite{Allen1989}.
Due to the $r^{-2}$ form, the direct summation over all interacting dipoles cannot converge with finite cutoff in real space.
Ewald summation divides this into three converged terms, as
\begin{equation}
\label{edipole}
E^{(\text{3D})}_{\text{dipole}}=E_{\text{r}}+E_{\text{k}}+E_{\text{corr}}.
\end{equation}
The real space term $E_{\text{r}}$ can quickly converge in real space, as
\begin{equation}
\label{dipole-r}
E_{\text{r}}=\frac{1}{2}\sum_{i,j=1}^{N}{\sum_{\left|\mathbf{n}\right|=0}^{\infty}}^\prime\left((\bm{\mu}_i\cdot\bm{\mu}_j)B(\mathbf{r}_{ij}+\mathbf{n})-(\bm{\mu}_i\cdot\mathbf{r}_{ij})(\bm{\mu}_j\cdot\mathbf{r}_{ij})C(\mathbf{r}_{ij}+\mathbf{n})\right),
\end{equation}
with
\begin{equation}
B(r)=\frac{\text{erfc}(\kappa r)}{r^3}+\frac{2\kappa}{\sqrt{\pi}}\frac{e^{-\kappa^2r^2}}{r^2},
\end{equation}
\begin{equation}
C(r)=\frac{3\text{erfc}(\kappa r)}{r^5}+\frac{2\kappa}{\sqrt{\pi}}\frac{(2\kappa^2r^2+3)}{r^2}\frac{e^{-\kappa^2r^2}}{r^2},
\end{equation}
where $\kappa$ is the Ewald parameter, and $\bm{\mu}_i$ is the dipole at site $i$, and $\mathbf{n}$ represents the lattice vector $\left\{n_1\bm{a}+n_2\bm{b}+n_3\bm{c}|\mathbf{n}=(n_1,n_2,n_3)\right\}$. $E_{\text{r}}$ sums over all pair $\left<i,j\right>$ and all integer vector $\mathbf{n}$, with the prime meaning excluding $i=j$ for $\left|\mathbf{n}\right|=0$. And the k-space term can quickly converge in kspace
\begin{equation}
\label{dipole-k}
E_{\text{k}}=\frac{1}{2}\sum_{i,j=1}^{N}\sum_{\mathbf{k}\neq0}\frac{4\pi}{k^2L^3}(\bm{\mu}_i\cdot\mathbf{k})(\bm{\mu}_j\cdot\mathbf{k})e^{-\frac{k^2}{4\kappa^2}}\cos(\mathbf{k}\cdot\mathbf{r}_{ij}),
\end{equation}
where  $\bm{k}$ labels the reciprocal lattice vector. And the correction term is
\begin{equation}
\label{dipole-corr}
E_{\text{corr}}=-\sum_i^N\frac{2\kappa^3}{3\sqrt{\pi}}\bm{\mu}_i^2+\frac{1}{2}\sum_{i,j=1}^N\frac{4\pi}{3L^3}\bm{\mu}_i\cdot\bm{\mu}_j.
\end{equation}
Note that $E^{(\text{3D})}_{\text{dipole}}$ is written in forms of $\bm{\mu}_i$ which might require DFT calculations. 
Actually, we can equivalently derive the polarization via FE mode magnitude $\bm{u}_i$.
Modern polarization theory tells that the changed polarization is linear to the displacement of the centers of Wannier function, indicating an access to calculate the polarization $\bm{\mu}_i$ of crystals,\cite{Resta1993} through
\begin{equation}
\label{borncharge}
\bm{\mu}_i=\frac{eZ_{\text{Born}}^*}{V}\mathbf{u}_i,
\end{equation}
where $Z_{\text{Born}}^*$ is the Born effective charge derived from first-principles calculations.
Substituting $\bm{\mu}_i$ with $\mathbf{u}_i$, we would rewrite the dipole-dipole interaction in a form of FE modes $\bm{u}_i$, as
\begin{equation}
\label{edipole}
E_{\text{dipole}}^{(\text{3D})}=\frac{1}{2}\sum_{i,j=1}^N\sum_{\alpha\beta}Q_{ij,\alpha\beta}u_{i_\alpha}u_{j_\beta},
\end{equation}
where the coefficients $Q_{ij,\alpha\beta}$ are calculated from Eq.~(\ref{dipole-r})-(\ref{dipole-corr}) once and for all, and are stored for latter $E_\text{dipole}$ calculations.
Ensuring efficiency and accuracy, we shall update these coefficients only when the lattice is largely distorted from the last calculated structure.
The $E_{\text{short}}$ is written in a similar form with Eq.~(\ref{edipole}), as
\begin{equation}
\label{eshort}
E_{\text{short}}=\frac{1}{2}\sum_{\left<i,j\right>}\sum_{\alpha\beta} J_{ij,\alpha\beta}u_{i\alpha}u_{j\beta},
\end{equation}
where $J_{ij,\alpha\beta}$ is the interaction coefficients. Unlike $E_{\text{dipole}}$, $\left<i,j\right>$ sums over atmost the 
third nearest neighbor (TNN. FNN and SNN for first and second nearest neighbor in the same way), while farther inter-site interaction has been cut off in condition dipole-dipole interactions have 
been excluded.
In fact, we determine this cutoff by watching the tendency of interaction magnitude towards farther sites. 
It shows a ratio of 1\ :\ 0.180\ :\ 0.014 by comparing the strongest magnitudes of the FNN, SNN, and TNN interaction parameters.
Considering that there are 6 FNNs, 12 SNNs, and 8 TNNs, the ratio of energy contributions comes to be 1\ :\ 0.360\ :\ 0.018.
This cutoff up to TNN is reasonable since it quickly converge and the ignored parts would bring slight contributions.
The independent parameters are $j_1-j_7$, labeling different interacting configurations which could be found in Ref.~[\onlinecite{Zhong1995}].
Then we look at the lattice strain part $E_2(\left\{\eta_l\right\},p_{\text{ext}})$ and its interaction with FE modes $E_3(\left\{\bm{u}_i\right\},\left\{\eta_l\right\})$.
\begin{equation}
E_2(\left\{\eta_l\right\},p_{\text{ext}})=E_{\text{strain}}(\left\{\eta_l\right\})+E_{\text{p}}(\left\{\eta_{1,2,3}\right\},p_{\text{ext}})
\end{equation}
\begin{equation}
E_3(\left\{\bm{u}_i\right\},\left\{\eta_l\right\})=E_{\text{int}}(\left\{\bm{u}_i\right\},\left\{\eta_l\right\})
\end{equation}
Lattice strain is crucial to describe structural properties of FE materials.
The strain contains homogeneous and inhomogeneous part, describing the averaged and localized structure distortion, respectively.
Without introducing lattice strain, BTO shows incorrect phase sequence as cubic to rhombohedral phase.\cite{Zhong1995}
When considering the controversial domain structures in SnTe films, inhomogeneous strain must be introduced.
This requires deepening studies, however, being off the main issues of this article.
Since the considered FE phase transition is more about the macroscopic properties, we take only the homogeneous part of the strain to simplify the problem.
The strain energy is
\begin{equation}
\label{estrain}
E_{\text{strain}}=\frac{N}{2}B_{11}(\eta_1^2+\eta_2^2+\eta_3^2)+NB_{12}(\eta_1\eta_2+\eta_2\eta_3+\eta_3\eta_1)+\frac{N}{2}B_{44}(\eta_4^2+\eta_5^2+\eta_6^2),
\end{equation}
where $B_{11}$, $B_{12}$, and $B_{44}$ are elastic coefficients (they relate to elastic constant $C_{11}$ via $B_{11}=C_{11}*V_{\text{cell}}$, etc.), and $N$ is the number of simulated cells. 
The external pressure $p_{\text{ext}}$ is here to consider the lattice mismatch with the substrates, further enabling investigation on the pressure-temperature phase diagrams. It interacts with strain via
\begin{equation}
\label{ep}
E_{\text{p}}=p_{\text{ext}}\cdot \Delta V=p_{\text{ext}}V(\eta_1+\eta_2+\eta_3).
\end{equation}
And the interaction between strain and on-site FE mode is
\begin{equation}
\label{eint}
E_{\text{int}}=\frac{1}{2}\sum_{i=1}^N\sum_{l\alpha\beta}B_{l\alpha\beta}\eta_l u_{i\alpha}u_{i\beta},
\end{equation}
where $B_{l\alpha\beta}$ is the coupling coefficients.
Due to cubic symmetry, we have written Eq.~(\ref{estrain}) and Eq.~(\ref{eint}) with only independent coefficients. 
Here, $B_{11}$, $B_{12}$, and $B_{44}$ are independent for the strain energy, and $B_{1xx}$, $B_{1yy}$, and $B_{4yz}$ are independent for the coupling energy of strain and FE modes.
Finally, combining all the terms in Eq.~(\ref{eself})(\ref{edipole})(\ref{eshort})(\ref{estrain})(\ref{ep})(\ref{eint}), we obtain the total energy as
\begin{equation}
\label{efinal}
E_{\text{3D-param}}^{(\text{3D})}=E_{\text{self}}+E_{\text{dipole}}^{(\text{3D})}+E_{\text{short}}+E_{\text{strain}}+E_{\text{p}}+E_{\text{int}},
\end{equation}
where all the parameters are derived from the first-principles calculations.
We shall use Eq.~(\ref{efinal}) latter to perform Monte Carlo simulations.
One finite-$T$ FE modes configuration on strained lattice is shown schematically in Fig.~\ref{figs1}.
We set a finite temperature $T$ and external pressure $p_{\text{ext}}$, allowing FE modes $\left\{\bm{u}_i,~ i=1,2,\cdots,N\right\}$, lattice strain $\left\{\eta_l,~l=1,2,\cdots,6\right\}$ varied to achieve thermal equilibrium.
Then we do statics on ${u}_{x,y,z}=\left<{u}_i\right>_{x,y,z}$ and $\left\{\eta_l,~l=1,2,\cdots,6\right\}$ to tell the properties of the system at finite-$T$.
The structural informations are given by the distorted lattice and displaced atomic positions.
Ferroelectric properties are described by the alignments of FE modes.
When they aligned uniformly to one direction, a FE phase is determined and the magnitude of polarization is given by Eq.~(\ref{borncharge}).
And when they aligned randomly with the statistical average to be zero, a PE phase is determined.

\subsection{2D case}
Coming to the 2D case, the effective Hamiltonian must be modified to restore the varied electronic structure.
Our main purpose is to build the bridge between bulk and thin films, upon which understand the abnormal behavior of $T_{\text{c}}$ in SnTe.
A natural choice is to retain the effective Hamiltonian of bulk and introduce some corrections to characterize the own properties of the films.
The local terms $E_{\text{self}}$ originate from the isolated on-site energy of the FE modes.
Thereby if the lattice constant and atomic position remain unchanged, $E_{\text{self}}$ of this site should remain unchanged no matter when its surrounding sites form 2D geometry or 3D geometry.
$E_{\text{dipole}}$ would receive changes upon going to films, however, mainly from the geometry changes (lattice contraction in out-of-plane direction and expansion in in-plane direction).
The picture of point dipoles at each site from macroscopic viewing remains unchanged.
By definition, the short-range terms $E_{\text{short}}$ are mostly affected.
The cut in out-of-plane direction leads to electrons redistribution, one might roughly consider, from original position interacting with out-of-plane neighboring sites to in-plane neighboring sites.
Therefore, we keep the the form of local terms $E_{\text{self}}$ unchanged, consider only geometry changes in dipole-dipole interaction terms (from $E_{\text{dipole}}^{(\text{3D})}$ to $E_{\text{dipole}}^{(\text{2D})}$), and count all the other surface effects in short range terms (keep $E_{\text{short}}$ unchanged and add corrections in $E_{\text{corr}}^{(\text{2D})}$).
The total of the energy in 2D geometry is written as:
\begin{equation}
E_{\text{3D-param}}^{(\text{2D})}\left(\{\mathbf{u}_i\},\eta,p\right)=E_{\text{self}}+E_{\text{dipole}}^{(\text{2D})}+E_{\text{short}}+E_{\text{strain}}+E_{\text{p}}+E_{\text{int}}+E_{\text{corr}}^{(\text{2D})},
\end{equation}
where only $E_{\text{dipole}}^{(\text{2D})}$ and $E_{\text{corr}}^{(\text{2D})}$ are different from the 3D case.
In order to satisfy the prerequisite that when increasing the film thickness the correction terms $E_{\text{corr}}^{(\text{2D})}$ must vanish spontaneously, we take the form of Ref.~[\onlinecite{Almahmoud2004}] and add the exponential decay coefficients to describe this layer dependency, as
\begin{equation}
\label{2D_corr}
E_{\text{corr}}^{(\text{2D})}\left(n_l\right)=\sum_{\substack{\alpha=\beta\\ \alpha=x,y}}\sum_{\substack{\left<i,j\right>\\j=i\pm\hat{\alpha}}} e^{-B\cdot n_l} A_{ij,\alpha\beta}u_{i\alpha}u_{j\beta},
\end{equation}
where $n_l$ labels the number of layers, $A_{ij,\alpha\beta}$ describes the short range interactions
(exclude the short part of dipole-dipole interactions) between neighboring sites $\left<i,j\right>$.
Meanwhile, EW2D instead of EW3D should be used to solve the $E_{\text{dipole}}$ terms in the Hamiltonian due to slab geometry.
The EW2D summation is also written in three parts
\begin{equation}
E^{(\text{2D})}_{\text{dipole}}=E_{\text{r}}+E_{\text{k}}+E_{\text{corr}},
\end{equation}
where $E_{\text{r}}$ is the same as bulk, but kspace terms $E_{\text{k}}$ and correction terms $E_{\text{corr}}$ are different from bulk, as
\begin{equation}
\begin{aligned}
E_{\text{k}}=&\frac{1}{2}\sum_{i,j=1}^N\sum_{\mathbf{k}\neq 0}\frac{\pi}{L^2}e^{i\mathbf{k}\cdot\bm{\rho}_{ij}}\\
&\left\{ (\bm{\mu}_i^{\rho}\cdot\mathbf{k})(\bm{\mu}_j^{\rho}\cdot\mathbf{k})D(z_{ij})-i[\mu_i^z(\bm{\mu}_j^{\rho}\cdot\mathbf{k})+\mu_j^z(\bm{\mu}_i^{\rho}\cdot\mathbf{k})]\frac{\partial D(z_{ij})}{\partial z_{ij}}-\mu_i^z\mu_j^z\frac{\partial^2 D(z_{ij})}{\partial z^2_{ij}}\right\},
\end{aligned}
\end{equation}
with
\begin{equation}
D(z_{ij})=\frac{1}{k}\left[e^{kz}\text{erfc}(\frac{k}{2\kappa}+\kappa z)+e^{-kz}\text{erfc}(\frac{k}{2\kappa}-\kappa z)\right],
\end{equation}
where superscript $\rho$ labels the in-plane components and $z$ labels the out-of-plane component. And
\begin{equation}
E_{\text{corr}}=\frac{2\kappa\sqrt{\pi}}{L^2}\sum_{i,j=1}^N\mu_i^z\mu_j^z\ e^{-\kappa^2z_{ij}^2}-\frac{2\kappa^3}{3\sqrt{\pi}}\sum_{i=1}^N\mu_i^2.
\end{equation}

\begin{figure}[b]
\includegraphics[width=0.4\linewidth]{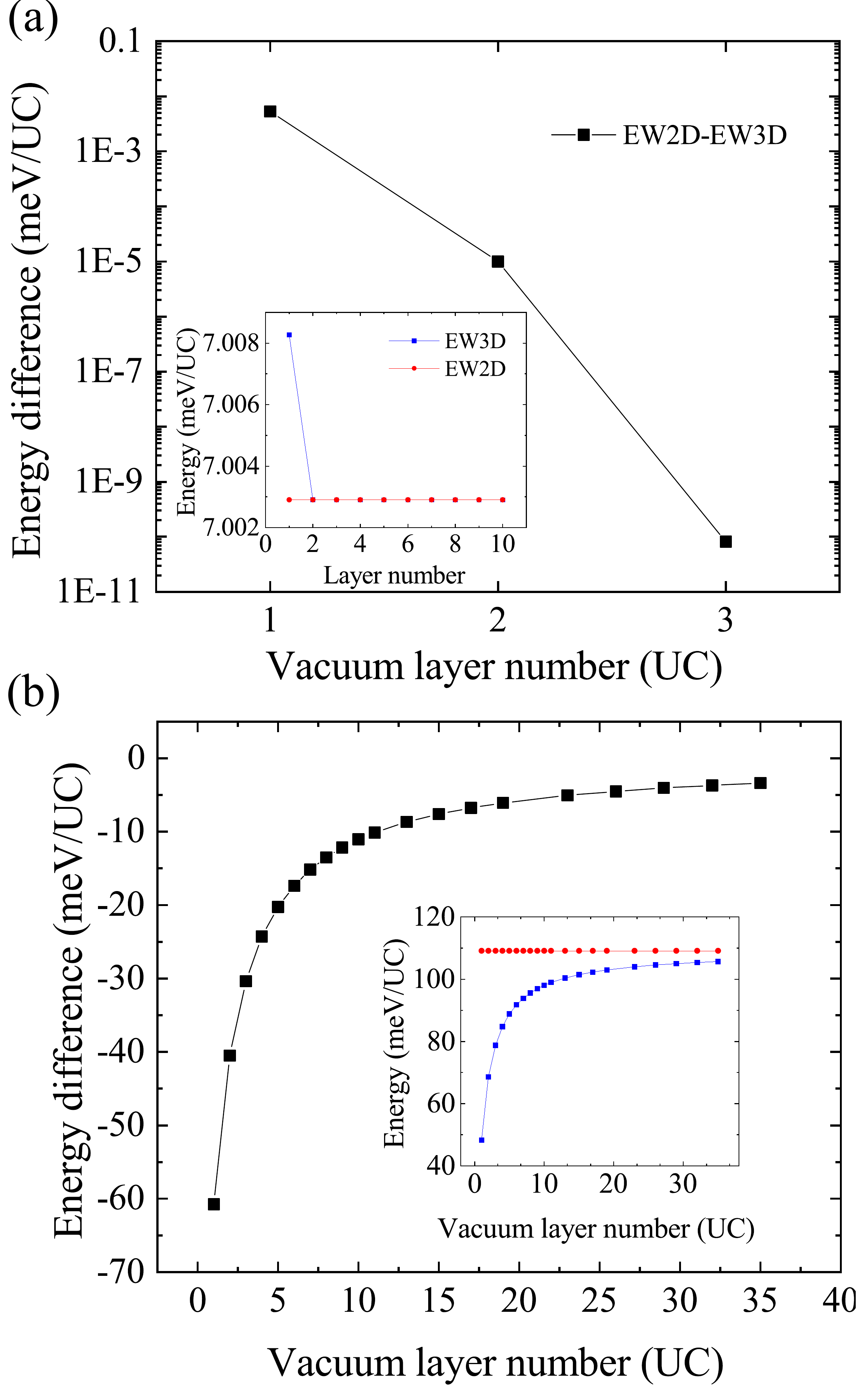}
\caption{\label{figs2} EW3D and EW2D results for $E_{\text{dipole}}$ calculations in film geometry. (a) shows the in-plane polarization case and (b) shows the out-of-plane polarization case. }
\end{figure}
In Fig.~\ref{figs2}, we show a test for calculating dipole-dipole interaction via two commonly used method, EW2D and EW3D. 
To derive the energy for a same system of $xy$-periodic and $z$-free (ppf periodic boundary condition) film, EW2D can be directly used while EW3D requires extra treatment. 
EW3D expand this film to a $xyz$-periodic structure with a few UC vacuum (ppp periodic boundary condition).
EW3D method will be fast iff the vacuum is not too large, but rough due to its unconsidered layer interaction errors.
When only deal with in-plane polarization, EW3D reaches convergence with EW2D upon 3UC vacuum.
However, EW3D shows slow convergence for out-of-plane polarization upon increasing vacuum slab thickness.
Considering this, we will use EW2D for 2D calculation to ensure our results.
\section{Computational details}
\subsection{DFT calculations}
\begin{figure}[b]
\includegraphics[width=0.8\linewidth]{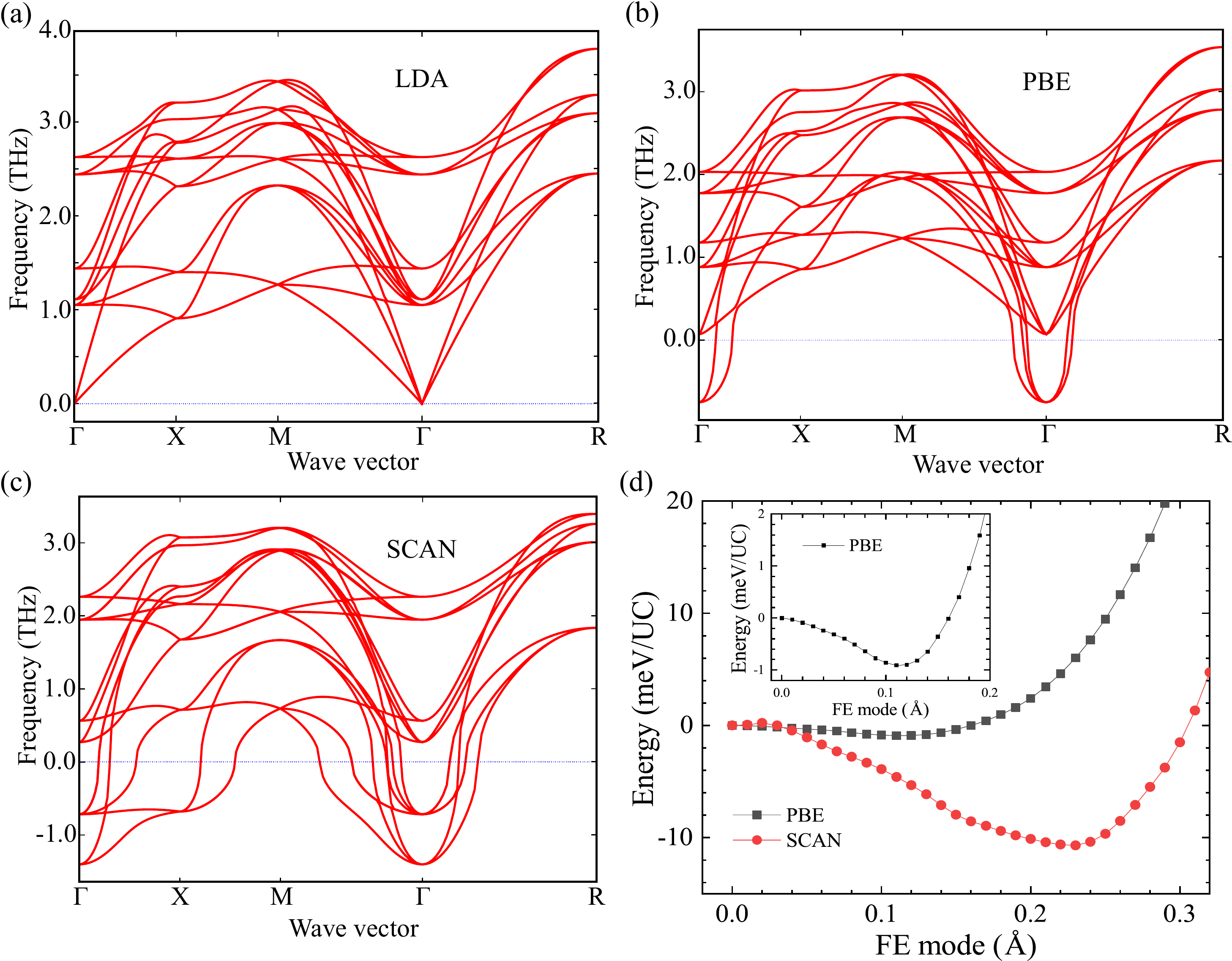}
\caption{\label{figs3} Phonon spectrum of SnTe via different functionals. (a)-(c) are LDA, PBE, and SCAN results, respectively\footnote{The non-zero acoustic mode in SCAN results might originate from some bugs of the program codes. We found this bug reoccurs in phonon calculations of perovskites, hexaferrites, and more materials. Since all the required informations from DFT calculations are the total energies, we trust in these energies.}. Note that non-analytical term correction (characterizing the LO-TO spliting) aren't included here. (d) the potential energy curves from DFT calculations using different functionals. Since LDA presents no soft mode, here we only show PBE results (in black curves) and SCAN results (in red curves). The atomic configurations have been displaced following the patterns of FE mode. We noticed that PBE results in a shallower well than SCAN.}
\end{figure}

For DFT calculations, we use the Vienna ab initio Simulation Package (VASP) with projector-augmented-wave (PAW) method.\cite{Kresse1996, Kresse1999}
The PAW valence electron configurations are $4d^{10}5s^25p^2$ for Sn and $5s^25p^4$ for Te.
The PAW energy cutoff is 850 eV for SCAN functional.
A $12\times12\times12$ k-point mesh turns out to yield converged results for the Bravais cell of bulk SnTe (a cubic structure with 4 Sn atoms and 4 Te atoms).
We use finite-displacement method for phonon calculations, as implemented in PHONOPY.\cite{Togo2015}
In Fig.~\ref{figs3}, we show the results with LDA,\cite{Ceperley1980} PBE,\cite{Perdew1996} and
SCAN functionals.\cite{Sun2015}
LDA yields no soft mode, while PBE and SCAN yield soft modes.
And the shear modes also turn to be a soft mode in SCAN, which might be crucial for group-IV monochalcogenides,\cite{Rabe1987a} and we shall count it in the shear strain terms of the Hamiltonian.
Potential energy curves following the displacement patterns of the FE mode are then calculated.
PBE curves process a shallower well than SCAN.
We think PBE might not be accurate enough to describe the electronic structure of SnTe.
Fei \textit{et al.} actually have derived non-polar structure for SnTe film using the PBE
functional (Fig.~S3(b) in the supplemental information of Ref.~[\onlinecite{Fei2016}]), in contrast
with experiment.\cite{Chang2016}
Hence, we use SCAN functionals for DFT calculations and later parameterizations of the effective
Hamiltonian.

\begin{table}[h]
\caption{Lattice constant by DFT calculations}
\begin{tabular}{p{5cm}p{2.5cm}<{\centering}p{2.5cm}<{\centering}p{2.5cm}<{\centering}p{2.5cm}<{\centering}}
\hline\hline
Functionals & LDA & PBE & SCAN & exp. \\\hline
Lattice constant(\AA) & 6.24 & 6.40 & 6.34 & 6.32 \\
\hline\hline
\end{tabular}
\end{table}

For film calculations, we also base on this cubic structure and build the supercell with a
vacuum of 19.03~\AA\text{} along $z$-direction (out-of-plane $\left[001\right]$ direction).
The $z$-grid of k-mesh is varied in associate with the length of $z$-direction.
As in experiments, the SnTe films interact weakly with the substrates (graphitized 6-H SiC covered
by mono-/bi-layer graphene) through van der Waals interactions which are often viewed as truly 2D materials.
Hence we apply free-standing films, retaining its intrinsic 2D features.
This also means the open circuit (OC) condition should be used for the following Monte Carlo (MC) simulations.
The van der Waals (vdW)-type interactions aren't introduced to avoid double counting, due to the
features of SCAN functionals.\cite{Sun2015}
Besides, since strain effects has been included in the Hamiltonian (see section I), we use the same
lattice constant as bulk for films.
In BaTiO$_3$ (BTO), the LDA functional is used for an easier comparison with published
results.\cite{Zhong1994,Zhong1995}
Concerning the fact that BTO has been systematically studied in these reference, and it is a system when conventional scaling law holds, we will mainly focus on SnTe in later discussions and resort to BTO only when comparison is needed.

\subsection{Parameters}
Here we list the parameters of the effective Hamiltonian for BTO and SnTe from above first-principles calculations.
\begin{table}[h]
\caption{{\label{BTO_param}}Parameters of the effective Hamiltonian for BTO. Energies are in hatrees in order to compare with Ref.~[\onlinecite{Zhong1995}].}
\begin{tabular}{ccccccc}
\hline\hline
Self & $\kappa_2$ & 0.0691 & $\alpha_4$ & 0.328 & $\gamma_4$ & -0.504\\\hline
Dipole & $\epsilon$ & 5.24 & $Z^*_{\text{Born}}$ & 10.26\\\hline
Short & $j_1$ & -0.0288 & $j_2$ & 0.0393 \\
& $j_3$ & 0.00979 & $j_4$ & -0.0106 & $j_5$ & 0.00580 \\
& $j_6$ & 0.00492 & $j_7$ & 0.00246 \\\hline
Strain & $B_{11}$ & 4.76 & $B_{12}$ & 1.62 & $B_{44}$ & 1.82 \\\hline
Inter & $B_{1xx}$ & -2.25 & $B_{1yy}$ & -0.171 & $B_{4yz}$ & -0.0888 \\\hline
2D corr & $A$ & 0.0282 & $B$ & -0.4958\\
\hline\hline
\end{tabular}
\end{table}

As a verification, we first independently obtain the parameters of BTO, which are compared with those of Vanderbilt.\cite{Zhong1995}
LDA functional is used here for BTO.
The parameters for BTO are listed in Table.~\ref{BTO_param}.
One could easily see all our parameters are close to those of Vanderbilt (Table.~II in Ref.~[\onlinecite{Zhong1995}]).

\begin{table}[h]
\caption{{\label{SnTe_param}}Parameters of the effective Hamiltonian for SnTe. Energies are in hatrees in associate with BTO.}
\begin{tabular}{ccccccc}
\hline\hline
Self & $\kappa_2$ & 0.0128 & $\alpha_4$ & 0.0140 & $\gamma_4$ & -0.00971\\\hline
Dipole & $\epsilon$ & 51.9 & $Z^*_{\text{Born}}$ & 19.9\\\hline
Short & $j_1$ & -0.00407 & $j_2$ & 0.000402 \\
& $j_3$ & 0.000128 & $j_4$ & -0.000731 & $j_5$ & 0.000457 \\
& $j_6$ & 0.0000582 & $j_7$ & 0.0000291 \\\hline
Strain & $B_{11}$ & 6.82 & $B_{12}$ & 0.0972 & $B_{44}$ & 1.09 \\\hline
Inter & $B_{1xx}$ & -0.264 & $B_{1yy}$ & -0.0270 & $B_{4yz}$ & -0.0165 \\\hline
2D corr & $A$ & -0.00722 & $B$ & -0.376\\
\hline\hline
\end{tabular}
\end{table}

The parameters for SnTe are listed in Table.~\ref{SnTe_param}.
By comparing the dimensionless parameters (strain related $B_{11}, B_{12}, B_{44}$) with BTO, we notice that SnTe tends to rhombohedral structure.
Lowered $B_{12}$ results in less competition between different direction of lattice strain, and lowered $B_{44}$ allows larger shear strain.
Since the polarization and size of unit cell are different for BTO and SnTe, direct comparisons of the magnitude of other parameters related with FE modes are meaningless.
A better choice is to compare their relative value for each material.\footnote{Another choice is to reduce all the parameters by polarization density. This bring a factor of $Z/L^3$ to the length dimension. Parameters are reduced by their own length dimension, \textit{e.g.} $\kappa_2$ has the power of two and $\gamma_4$ has the power of four.}
We shall see for short-range interaction parameter, the numerical relations are similar for SnTe and BTO.
However, the $\alpha_4$ and $\gamma_4$ are lower and $B_{4yz}$ are larger in SnTe than BTO.
Since competition between different direction is lowered, the tendency to rhombohedral phase in SnTe should be reserved.
Actually, our simulations tell that in BTO the polarizations of different directions appear in sequence, exhibiting four phases as cubic to tetragonal to orthogonal to rhombohedral.
In SnTe bulk, the polarization appear simultaneously in all three directions, exhibiting only two phases as cubic to rhombohedral.

\subsection{Monte Carlo calculations}
We use the aforementioned Hamiltonian with parameters derived in last subsection.
Due to that most energy terms of the Hamiltonian are localized expect the dipole-dipole interaction, we use the single flip algorithm.
That is, each Monte Carlo sweep (MCS) consists of a series of trial moves of the FE modes on each site and the six components of the homogeneous strain, in which homogeneous strain components take 20-100 trial moves repeatedly in one MCS.
The step sizes are adapted to control the accept ratio in range of 20\%-30\%.
For each simulation configuration (cell, temperature, and external pressure), we run at least 200,000 MCSs, in which first 150,000 MCSs are used to ensure thermal equilibrium and last 50,000 MCSs are used for statistics.

\begin{figure}[h]
\includegraphics[width=0.5\linewidth]{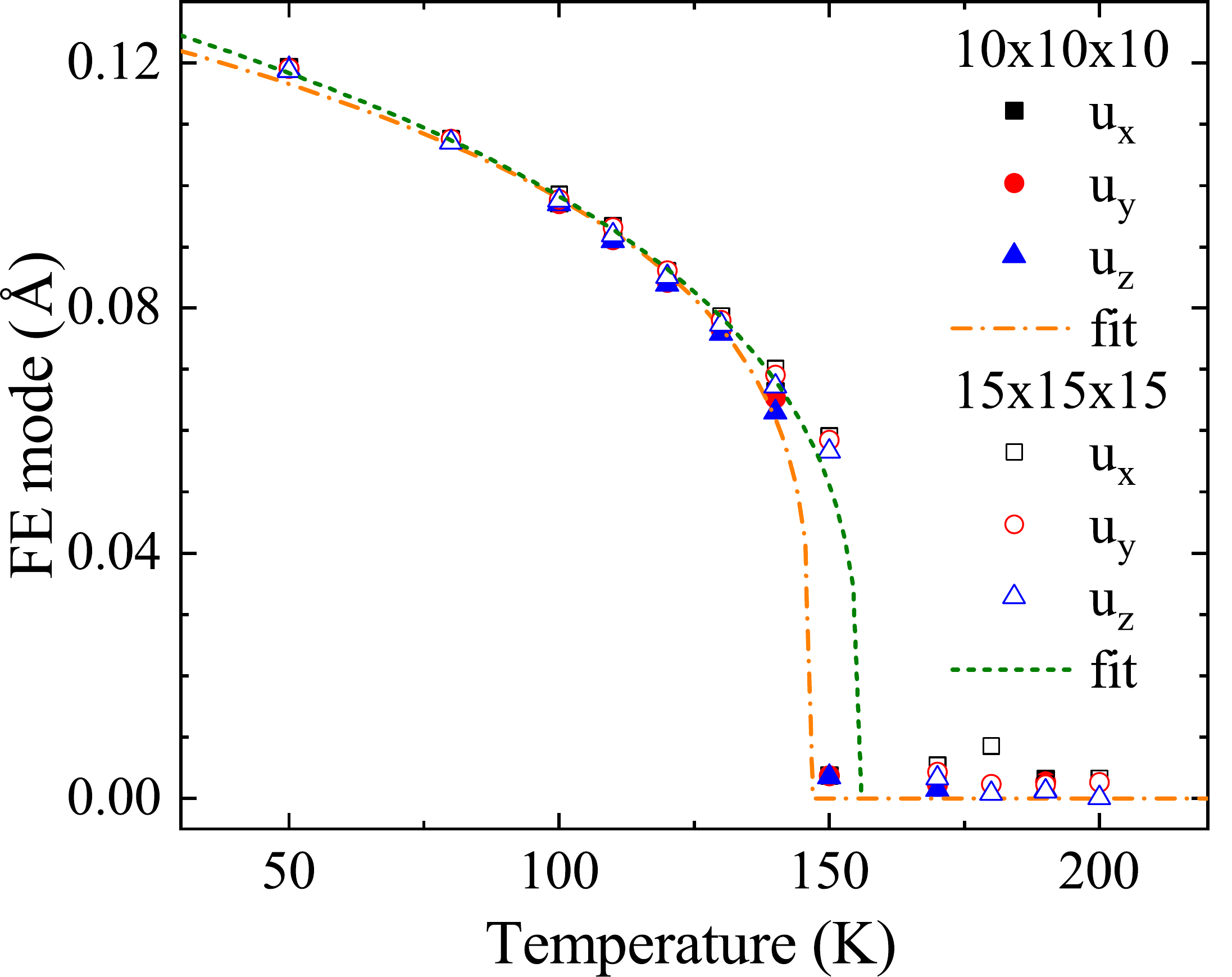}
\caption{\label{figs4} Convergence test for the simulation cell. $10 \times 10 \times 10$ supercell and $15\times15\times15$ supercell are in coincidence at low temperature and show a few deviation near the critical point. $10\times10\times10$ cell shows $\sim$10 K difference from $15\times15\times15$ cell. In fact, this range of error is far lower than the difference between bulk and film (several hundreds of Kelvins), or among films of different layers (several tens to hundreds of Kelvins). Considering 4 times atoms and at least 16 times computation loads with $15\times15\times15$ cell, we adopt $10\times10\times10$ cell for latter simulations.}
\end{figure}

The long-range nature of ferroelectricity requires large-scale simulations.
Here we take the simulation cells as $10\times 10\times 10$ periodic supercell for bulk (corresponding to a cell with $a=63.2$\AA\ containing 8,000 atoms, which results in unaffordable computation load for DFT calculations. That's also why we use a model Hamiltonian method instead of \textit{ab initio} molecular dynamics), and $10\times10\times N$ $xy$-periodic $z$-free slab for $N$-UC films.
The convergence test for this supercell has been shown in Fig.~\ref{figs4}.
The errors from the cell sizes are controlled to a few Kelvin, which are accurate enough to tackle the issues in our manuscript, telling the thickness dependency of Curie temperature.
\newpage
\section{Pressure compensation}
The lattice strain has crucial influence in FE phase transitions.
Due to insufficiently accurate electronic structure of DFT, transition temperature might be far from the correct value by using the DFT optimized geometry.
An exerted pressure is used here to compensate this error.
We determine this compensation pressure by comparing with the experimental measurements.
As shown in Fig.~\ref{figs5}, we calculate the phase transitions of bulk SnTe at different external pressures.
We found $\sim$2GPa gives reasonable $T_{\text{c}}$ for bulk SnTe [Fig.~\ref{figs5}].
In the same way, the compensation pressure for 1-10UC films are all set as 1GPa, difference from bulk being used to count for the substrates strain.
This is in agreement with experiment by Chang \textit{et al.}, which applied substrates with lattice constant slightly larger than SnTe films.
We don't seek for a more precise value for this compensation pressure due to the effective Hamiltonian method might have as a few ten Kelvins as systematic errors.
Besides, this compensation pressure is reasonable in its magnitude.
That is, SCAN slightly overestimate the lattice constant, requiring the compensation pressure to be positive and small.
LDA, however, underestimate the lattice constant, requiring the compensation pressure to be negative and larger.
This is true in our results, since BTO with LDA uses a -8GPa compensation pressure\footnote{In Ref.~[\onlinecite{Zhong1995}], Vanderbilt use -4.8GPa compensation pressure in order to fit all the three transition temperature (namely C-T, T-O, O-R) in BTO. This leads to a $\sim$100K divergence with experiments for Curie temperature. Here we use -8Gpa to better fit the Curie temperature (namely C-T).}, and SnTe with SCAN using a 2GPa compensation pressure for bulk.
To emphasize, when studying the thickness dependency of $T_{\text{c}}$ in films, threats from the large extrinsic factor of lattice strain can be ruled out by setting a same compensation pressure in our method.
However, the other model methods such as Landau's phenomenological model, $\phi^4$ model, etc. cannot guarantee this, since strain terms are not considered and the parameterizations are done in one specific structure.
One could see from the strain-temperature phase diagram in Ref.~[\onlinecite{Fei2016}], $T_{\text{c}}$ for 1UC SnSe varies from 64K to 640K among different strains.
This uncertainty actually hinder the exploration of the thickness dependency of $T_{\text{c}}$.
In our method, the built-in strain-pressure relation makes it possible to exclude extrinsic strain effect and question the intrinsic size effects of SnTe's abnormalities.
Using a same external pressure, we have show the clearly different tendency upon going from bulk to thin films in SnTe [Fig.~3(b) in our manuscript].

\section{Bulk results and P-T phase diagram}
\begin{figure}[ht]
\includegraphics[width=1.0\linewidth]{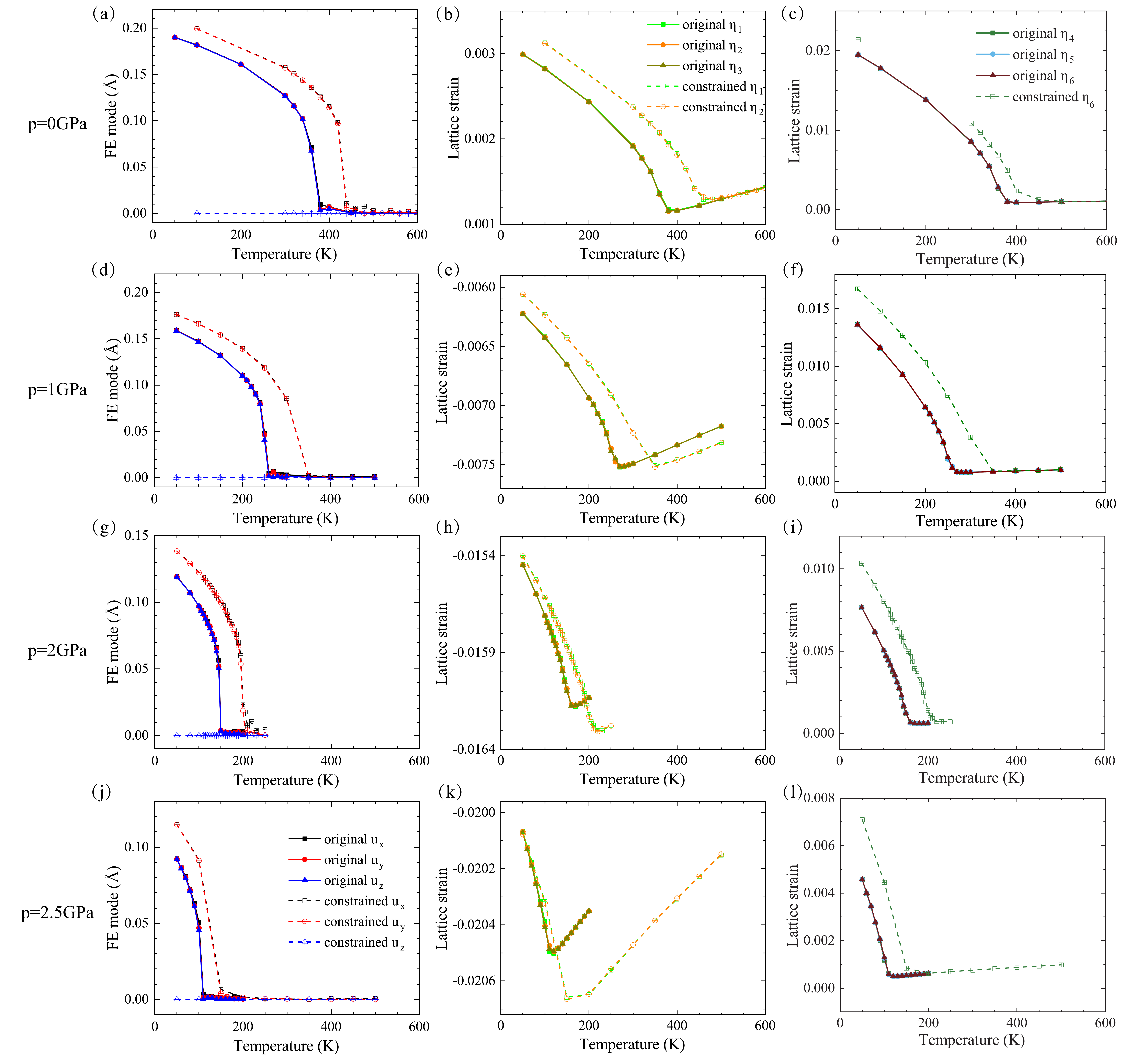}
\caption{\label{figs5} The microscopic details of the phase transitions in bulk SnTe under different pressures. Here we show the results of both original bulk (spontaneous phase transition) and the constrained bulk (constrain $u_z=0$ and $\eta_3=\eta_4=\eta_5=0$, we shall discuss later in section VI). (a)(d)(g)(j) shows the magnitudes of FE modes, (b)(e)(h)(k) and (c)(f)(i)(k) shows the magnitudes of diagonal and non-diagonal lattice strain, respectively.}
\end{figure}

\begin{figure}[h]
\includegraphics[width=0.6\linewidth]{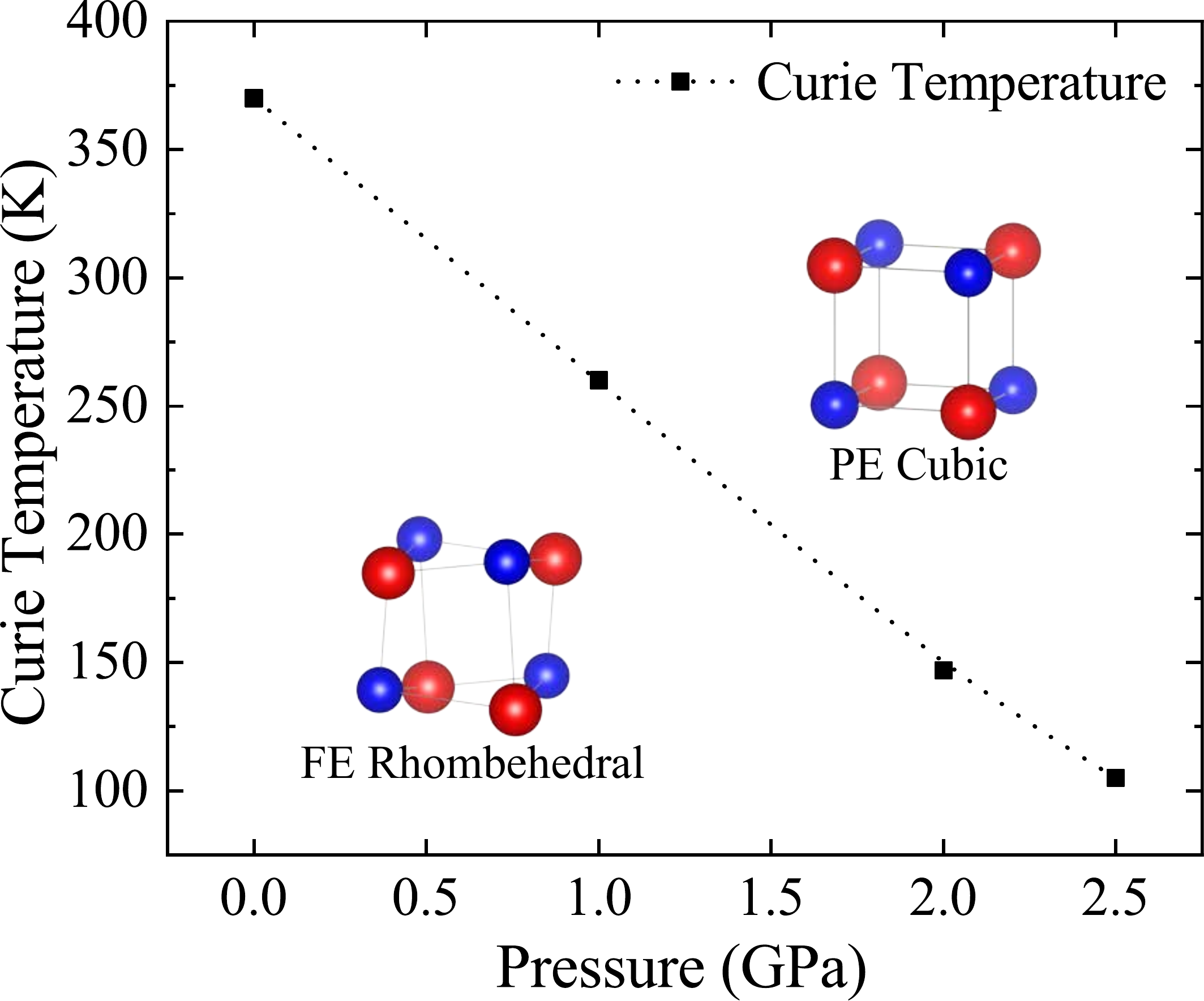}
\caption{\label{figs6} The pressure-temperature phase diagram of bulk SnTe. The compensation pressure we determined is 2GPa for bulk SnTe, which should be shifted when compring with the experiments. This leads to Curie temperature $T_{\text{c}}=147K$ corresponding to experiments at 0GPa, where $\sim$50K is left to count for the defect effects.}
\end{figure}

Here we process microscopic details of the phase transitions in SnTe bulk.
FE modes and lattice strain are used to determine the Curie temperature.
Taking Fig.~\ref{figs5}(g) as an example, we give the $T_{\text{c}}$ equals 147K for bulk with external pressure 2Gpa.
Above 147 K, the diagonal part of the strain tensor is the 
same and the average of all three components of the FE modes $\left<u\right>_{\alpha=x,y,z}$ are close to zero so as
the non-diagonal part of strain tensor $\left<\eta\right>_{i=4,5,6}$ [Fig.~\ref{figs5}(h) and \ref{figs5}(i)].
At 147~K, the magnitudes of the FE modes suddenly jump and then gradually saturate.
This divergence and the equivalence between the polarization in $x$, $y$, and $z$ directions indicate
that a rhombohedral FE phase is formed with polarization along $\left[111\right]$.
A more accurate determination of $T_{\text{c}}$ can be done by fitting the critical behavior of polarization, \textit{i.e.} FE modes (latter shown in Eq.~(\ref{fiteq})).
FE modes are viewed as the order parameters in FE phase transition.
We can convert it from $\mathrm{\AA}$ to $\mathrm{C} / \mathrm{m}^2$ via
\begin{equation}
\mathbf{P}=\frac{eZ^*_{\text{Born}}}{V}\left<\mathbf{u}\right>=1.26\left<\mathbf{u}\right>\frac{\mathrm{C}/\mathrm{m}^2}{\mathrm{\AA}}.
\end{equation}
Since structural phase transition occurs simultaneously, lattice strain could also be used to characterize the phase transitions.
\newpage
Utilizing above informations at different external pressures, we further derive the $P$-$T$ phase diagram of bulk SnTe, as shown in Fig.~\ref{figs6}.
In the pressure range we calculated, cubic PE phase and rhombohedral FE phase are observed.
Increasing the external pressure, the PE phase will take over the FE region, which is the same as previous studies in BTO.\cite{Zhong1995}
More phases might appear in much higher pressure, however, the above parameters are not appropriate to do these simulations.
This is because our effective Hamiltonian take the Taylor series and the parameters have the predictive power of first-principles if and only if this expansion is valid.
Much higher pressure breaks this expansion and a new reference structure along with new parameters are required.

\section{Film results}
\begin{figure}[t]
\includegraphics[width=0.7\linewidth]{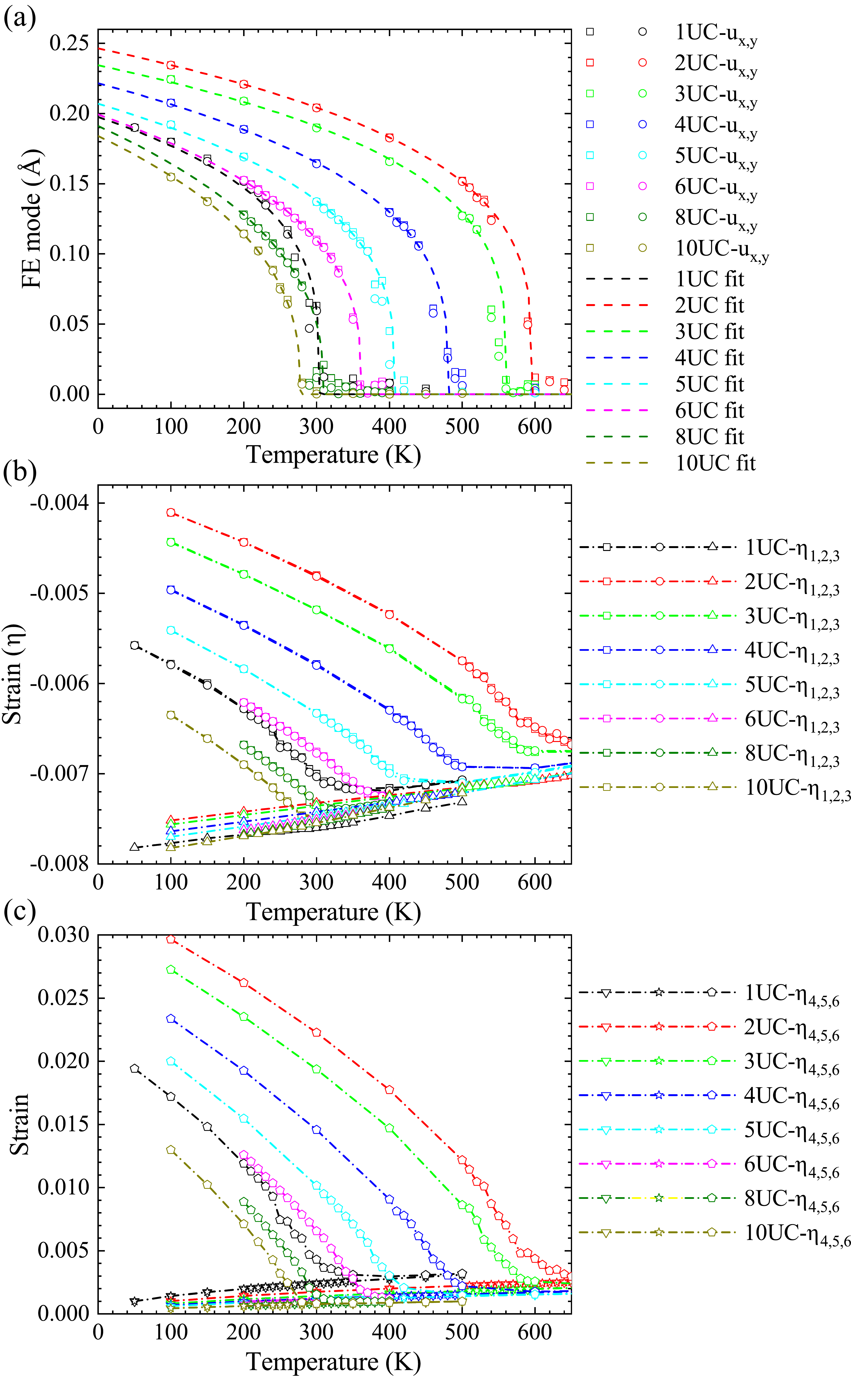}
\caption{\label{figs7} The microscopic details of the phase transitions in SnTe films with different layer numbers. (a) shows the magnitudes of the FE modes. The $u_z$ components in this range of thickness are nearly zero so that we don't show them for simplicity; (b)(c) shows the diagonal part and non-diagonal part of the lattice strain, respectively.}
\end{figure}

Here we process microscopic details of the phase transitions in SnTe films with different thicknesses, in Fig.~\ref{figs7}.
The experiments measured the distortion angle instead of the polarization.
In fact, the distortion angle corresponds to the non-diagonal strain [Fig.~\ref{figs7}(c)], especially $\eta_{6}$ here.
Their relation is
\begin{equation}
\Delta \alpha=90^\circ-\mathrm{arccos}(2\eta_{xy})=\mathrm{arcsin}(\eta_{6}).
\end{equation}
Since the distortion angle is small, above equation can be rewritten as
\begin{equation}
\Delta \alpha=\eta_{6}\cdot\frac{180^{\circ}}{\pi}.
\end{equation}
Our simulation gives the distortion angle for 1UC $\Delta \alpha_{\text{1UC}}\sim1.2^\circ$ extrapolated to $T=4K$, in agreement with experimental value $\sim1.4^\circ$.
Besides, we also use the following equation to fit the critical behavior of FE modes [short dash curves in Fig.~\ref{figs7}(a)]:
\begin{equation}
\label{fiteq}
u(T)=\begin{cases}
e^A(T_{\text{c}}-T)^\beta, & T<T_{\text{c}};\\
0, & T>T_{\text{c}}.
\end{cases}
\end{equation}
To be noted that, we fit the mean square root of the FE modes (for film, that is $u=\sqrt{(u_x^2+u_y^2)/2}$) instead of the total polarization.
This gives the critical indexes in consistent with the experimental value 0.33$\pm$0.05 for 1-4UC.\cite{Chang2016}

\begin{table}[htbp]
  \centering
  \caption{\label{table4} Curie temperature fit for films with different thicknesses.}
    \begin{tabular}{cp{2.5cm}<{\centering}p{2.5cm}<{\centering}p{2.5cm}<{\centering}}
    \hline\hline
    Layernumber   & \multicolumn{1}{c}{$T_{\text{c}}/K$} & \multicolumn{1}{c}{A} & \multicolumn{1}{c}{$\beta$} \\\hline
    1     & 304   & -3.198 & 0.2758 \\
    2     & 597   & -3.116 & 0.2684 \\
    3     & 562   & -3.157 & 0.2695 \\
    4     & 482   & -3.375 & 0.3024 \\
    5     & 408   & -3.410 & 0.3052 \\
    6     & 362   & -3.598 & 0.3370 \\
    8     & 310   & -3.884 & 0.3886 \\
    10    & 278   & -3.817 & 0.3774 \\
    \hline\hline
    \end{tabular}%
\end{table}%

\section{Constrained bulk simulation}
We process a numerical experiment for bulk SnTe.
Spontaneously, SnTe bulk has only two phases, the PE cubic phase and the FE rhombohedral phase.
However, the ultra-thin films seem more close to an orthogonal phase, which is absent in bulk.
By the convenience of effective Hamiltonian method, we can prepare a constrained bulk (c-bulk).
In this c-bulk, the polarization along $z$-direction has been constrained to zero, as well as other $z$-related components of the lattice strain, namely $\eta_3, \eta_4,$ and $\eta_5$.
Upon decreasing the thickness, the order parameters characterizing the FE phase transition are unchanged.
If we keep the form of interactions also unchanged, we shall find scaling law valid for this c-bulk.
We verify this in the main article [Fig.~3(b) olive curves in the manuscript].
Besides, since phase O has higher rank in the C-T-O-R transition sequence, we observe this c-bulk process higher $T_{\text{c}}$ than the original bulk (Fig.~\ref{figs5}).
We have showed in the main article, that the $T_c$ of films without correction terms are approaching c-bulk instead of original bulk.
Microscopic details of the c-bulk results are merged in Fig.~\ref{figs5}.

\newpage
\begin{figure}[t]
\includegraphics[width=0.6\linewidth]{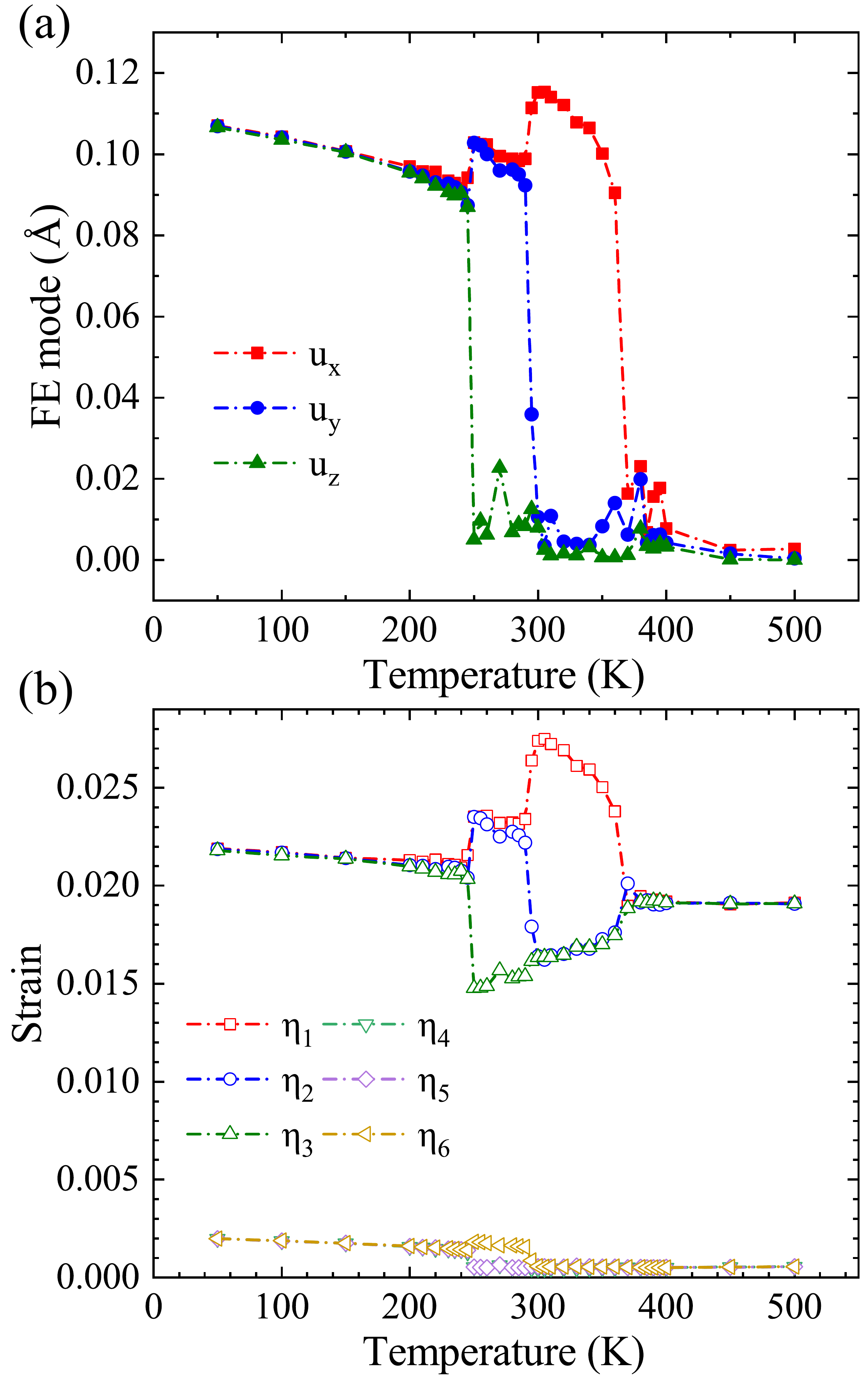}
\caption{\label{figs8} The pressure-temperature phase diagram of bulk BTO. Our results is close to Ref.~[\onlinecite{Zhong1995}]. We obtain C-T-O-R phase transition with reasonable transition temperatures.}
\end{figure}

%

\newpage